\documentclass[sn-nature]{sn-jnl}

\usepackage{graphicx}%
\usepackage{multirow}%
\usepackage{amsmath,amssymb,amsfonts}%
\usepackage{amsthm}%
\usepackage{mathrsfs}%
\usepackage[title]{appendix}%
\usepackage{xcolor}%
\usepackage{textcomp}%
\usepackage{manyfoot}%
\usepackage{booktabs}%
\usepackage{algorithm}%
\usepackage{algorithmicx}%
\usepackage{algpseudocode}%
\usepackage{listings}%
\usepackage[normalem]{ulem}

\newcommand\aj{{Astron.~J.}}
\newcommand\apj{{Astrophys.~J.}}
\newcommand\apjl{{Astrophys.~J.~Lett.}}
\newcommand\apjs{{Astrophys.~J.~Suppl.}}

\newcommand\aap{{Astron.~Astrophys.}}
\newcommand\mnras{{Mon.~Not.~R.~Astron.~Soc.}}
\newcommand\nat{{Nature}}
\newcommand\araa{{Ann.~Rev.~Astron.~Astrophys.}}

\newcommand\pasp{{Publ. Astron. Soc. Pac.}}

\newcommand{\noop}[1]{}

\begin{document}

\title[Article Title]{Evidence for Morning-to-Evening Limb Asymmetry on the Cool Low-density Exoplanet WASP-107b}

\author*[1]{\fnm{Matthew M.} \sur{Murphy}}\email{mmmurphy@arizona.edu} 
\author[2]{\fnm{Thomas G.} \sur{Beatty}} 
\author[1]{\fnm{Everett} \sur{Schlawin}} 
\author[3,4]{\fnm{Taylor J.} \sur{Bell}} 
\author[5]{\fnm{Michael R.} \sur{Line}} 
\author[4]{\fnm{Thomas P.} \sur{Greene}} 
\author[6]{\fnm{Vivien} \sur{Parmentier}} 
\author[7]{\fnm{Emily} \sur{Rauscher}} 
\author[5]{\fnm{Luis} \sur{Welbanks}} 
\author[8]{\fnm{Jonathan J.} \sur{Fortney}} 
\author[1]{\fnm{Marcia} \sur{Rieke}} 

\affil*[1]{\orgdiv{Steward Observatory}, \orgname{University of Arizona}, \orgaddress{\street{Cherry Ave.}, \city{Tucson}, \postcode{85705}, \state{AZ}, \country{USA}}}

\affil[2]{\orgdiv{Department of Astronomy}, \orgname{University of Wisconsin--Madison}, \orgaddress{\city{Madison}, \state{WI}, \country{USA}}}

\affil[3]{\orgdiv{Bay Area Environmental Research Institute}, \orgname{NASA's Ames Research Center}, \orgaddress{\city{Moffet Field}, \state{CA}, \country{USA}}}

\affil[4]{\orgdiv{Space Science and Astrobiology Division}, \orgname{NASA's Ames Research Center}, \orgaddress{\city{Moffet Field}, \state{CA}, \country{USA}}}

\affil[5]{\orgdiv{School of Earth and Space Exploration}, \orgname{Arizona State University}, \orgaddress{\city{Tempe}, \state{AZ}, \country{USA}}}

\affil[6]{\orgdiv{Université Côte d’Azur}, \orgname{Observatoire de la Côte d’Azur}, \orgaddress{\city{CNRS}, \state{Laboratoire Lagrange}, \country{France}}}

\affil[7]{\orgdiv{Department of Astronomy}, \orgname{University of Michigan}, \orgaddress{\city{Ann Arbor}, \state{MI}, \country{USA}}}

\affil[8]{\orgdiv{Department of Astronomy and Astrophysics}, \orgname{University of California Santa Cruz}, \orgaddress{\city{Santa Cruz}, \state{CA}, \country{USA}}}



\abstract{
The atmospheric properties of hot exoplanets are expected to be different between the morning and the evening limb due to global atmospheric circulation. Ground-based observations at high spectral resolution have detected this limb asymmetry in several ultra-hot ($>$2000 K) exoplanets, but the prevalence of the phenomenon in the broader exoplanetary population remains unexplored. Here we use JWST/NIRCam transmission spectra between 2.5 and 4.0 µm to find evidence of limb asymmetry on exoplanet WASP-107 b. With its equilibrium temperature of 770 K and low density of 0.126 gm c$^{-3}$, WASP-107 b probes a very different regime compared to ultra-hot giant planets and was not expected to exhibit substantial spatial heterogeneity according to atmospheric models. We infer instead a morning-evening temperature difference on the order of 100 K with a hotter evening limb. Further observations on other cooler exoplanets are needed to determine whether WASP-107 b is an outlier or the models underestimate the presence of limb asymmetry in exoplanets.
}

\maketitle

\section*{Main Text}

The WASP-107 system consists of an active K6 star \cite{mocnik2017_wasp107b, dai2017_wasp107b, hejazi2023_wasp107} orbited by at least two planets. First discovered in 2017 \cite{anderson2017_wasp107b}, the inner planet WASP-107b is among the lowest bulk-density exoplanets known to date. With a mass comparable to Neptune (M$_p$ = 30.5 M$_\oplus$ \cite{piaulet2021_wasp107b}) but a radius comparable to Jupiter (R$_p$ = 11.0 R$_\oplus$ \cite{anderson2017_wasp107b}), its bulk density of 0.126 g/cm$^3$ is roughly one-fifth that of Saturn, the lowest density planet in our Solar System. WASP-107b's density suggests it is a gaseous planet with a substantial atmosphere of primarily hydrogen and helium \cite{anderson2017_wasp107b, piaulet2021_wasp107b}. Further, such low density implies that WASP-107b has a uniquely large atmospheric scale height, which enhances the strength of molecular absorption features and makes it a favorable target for transmission spectroscopy. 

A transmission spectrum is measured from starlight that passes through the planet's atmosphere at its terminators, which are the dividing lines between the planet's day- and nightside. In the transiting geometry, the planetary limbs encompass its terminators. Tidally synchronized planets have perpetual day- and nightsides, leading to day-night irradiation contrast and global atmospheric circulation. Limb heterogeneity develops as this circulation advects heat in one direction, driving a temperature difference between the evening terminator, which is downstream of the heat flow, and the morning terminator, which is upstream \cite{kataria2016_gcmgrid, line2016_LA, schwartz2017_hotjupiteroffsets}. This heterogeneity is generally expected to be strongest on tidally synchronized planets with temperatures between $\sim$1200 - 2100~K, with cooler planets having homogeneous limbs \cite{kataria2016_gcmgrid, line2016_LA, powell2019_LAmodels}.

Several models have been used to simulate the signature of limb asymmetry in transmission \cite{line2016_LA, powell2018_cloudasym, powell2019_LAmodels}. In short, a difference in limb temperatures imparted by atmospheric circulation primarily causes a difference in scale height (via temperature) and cloud properties, which together alter the wavelength-dependent transmission through each limb. For example, with clouds, this temperature difference can be sufficient that species are gaseous in the evening limb, but have condensed in the morning limb. If neglected, limb asymmetry can lead to biased retrievals of atmospheric properties from low-resolution observations, including JWST \citep[c.f. \cite{macdonald2020_retrievalbias}]{line2016_LA, feng2016_retrievalbias, caldas2019_retrievalbias, welbanks2022_retrievalbias}. Therefore, observationally characterizing limb asymmetry is vital to ensuring that our insights to exoplanet atmospheres are bias-free and benchmarking models of atmospheric circulation and cloud formation. As mentioned, atmospheric models typically predict limb properties to homogenize below $\sim$1200 K, but this has not been observationally confirmed. At an equilibrium temperature of 770 K, WASP-107b lies in this cooler regime, and its large atmospheric scale height makes it a uniquely favorable target for investigating limb asymmetry at these temperatures. 

\section{Results}\label{sec:results}

\subsection{Observations} \label{subsec:results_observations}

\begin{figure}[h!]
    \centering
    \includegraphics{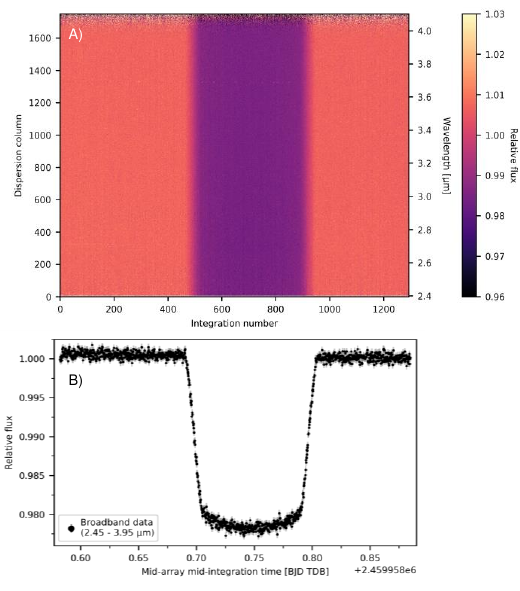}
    \caption{\textbf{Dynamic spectrum and broad-band light curve from our JWST/NIRCam F322W2 transit observation of WASP-107b.} Panel A shows our spectroscopic transit data at full spectral and temporal resolution, with wavelength on the y-axis and time on the x-axis in terms of the sequence of integrations. Each column is normalized by its median value, so the color-coding illustrates the corresponding relative flux values. In our analysis, we only use dispersion columns 56 - 1594, or wavelengths from 2.45 - 3.95~$\mu$m, in order to avoid excess noise visible on either edge of the detector. Panel B shows our broad-band light curve integrated over those wavelengths, with no additional light curve detrending done. Error bars represent the standard deviations (1-$\sigma$). Our variance is dominated by photon noise and follows a Poisson distribution. This light curve displays no strong non-linear systematic trends or starspot crossing events.}
    \label{fig:river_and_broadband}
\end{figure}

We observed one transit of WASP-107b on January 14th, 2023 using JWST/NIRCam as part of the MANATEE NIRCam GTO program (Obs.8, JWST-GTO-1185, ref.\cite{schlawin2018_manatee}). We simultaneously used NIRCam's F210M filter in the short-wavelength channel and the F322W2 grism in the long-wavelength channel \cite{NIRCamINSTRUMENT_Greene2010, NIRCam_inflightpaper}. For both, this observation was a continuous time-series of 1,293 integrations at 20.22-second cadence using the \texttt{BRIGHT2} readout pattern on the \texttt{SUBGRISM256} subarray. We reduced these data using three independent pipelines: \texttt{Eureka!} \cite{EUREKAPIPELINE}, \texttt{tshirt} (\url{https://github.com/eas342/tshirt}), and \texttt{Pegasus} \cite{beatty2024_gj3470b}. Each pipeline produced consistent results, discussed in the Methods, validating our results against systematic biases from the particular reduction method used. We focus primarily on the results from \texttt{tshirt} due to its more effective 1/f noise correction \cite{schlawin2020jwstNoiseFloorI}. We produced a band-integrated light curve at 2.093 $\pm$ 0.102~$\mu$m from F210M, as well as a band-integrated (2.45 -- 3.95~$\mu$m) light curve and thirty (R$\sim$64) spectroscopic light curves from F322W2. Our F322W2 dynamic spectrum and broad-band light curve are shown in Figure~\ref{fig:river_and_broadband}.

We also collected new and archival transit and radial velocity observations of WASP-107b using other instruments over a wide wavelength range. Uncertainty in the planet's orbital parameters, particularly the time of conjunction, can bias the signal of limb asymmetry \cite{line2016_LA, vonparis2016_LA, Espinoza2021_catwoman2}. To combat this, we use these ancillary data to precisely measure WASP-107b's orbit before our analysis of the JWST/NIRCam data.
These include radial velocity measurements with the CORALIE and Keck I/HIRES spectrographs collected by refs.\cite{anderson2017_wasp107b, piaulet2021_wasp107b}. Also, we downloaded public data covering three transits with the Transiting Exoplanet Survey Satellite, and one transit at 4.5~$\mu$m with the Infrared Array Camera on the Spitzer Space Telescope (Program 13052, PI: M. Werner). We observed one new transit in the SDSS i-band using the Goodman Spectrograph \cite{SOARGOODMANINSTRUMENT} on the Southern Astrophysical Research Telescope. Lastly, we obtained the broad-band (5 -- 12~$\mu$m) light curve of one transit using JWST's Mid-Infrared Instrument in Low-Resolution Spectroscopy mode (MIRI/LRS), observed in JWST program 1280 and given to us by PI P.O. Lagage. We describe the reduction and fitting of these data in the Methods, with particular focus on our derivation of WASP-107b's time of conjunction in Section~\ref{subsec:methods_tcfit}. The results are tabulated in Extended Data Table~1. With these data, we measured WASP-107b's time of conjunction to a 1-$\sigma$ precision of 0.7 seconds.

\subsection{Evidence for limb asymmetry} \label{subsec:results_LA}

\subsubsection{Limb transmission spectra}

We fit the spectroscopic light curves with asymmetric-limb transit models using the modeling package \texttt{catwoman} \cite{Espinoza2021_catwoman2, catwoman1}. Described further in Methods Section~\ref{subsec:fitting_asymlimb}, we fit for the planet's evening and morning limb planet-star radius ratios at each wavelength, and derived transmission spectra for each limb. We assume the planet's rotation axis is aligned to its orbit, discussed further in Methods Section~\ref{subsec:methods_obliquity}. We left the time of conjunction ($t_c$) as a free parameter with a Bayesian prior set to its posterior distribution from fitting our ancillary observations, and $t_c$ did not vary from this prior. The best-fit spectra are shown in Figure~\ref{fig:limbspectra} and tabulated in Extended Data Table~2.

We find that WASP-107b's morning and evening limbs have significantly ($>$99\% confidence) different transmission spectra, as seen in Figure~\ref{fig:limbspectra}. Across our bandpass, WASP-107b's evening limb is generally larger than its morning limb. The evening limb spectrum is $\sim$250 ppm higher in their pseudo-continuum regions where the limb depths differ by $\sim$2-$\sigma$, and the corresponding radii differ by roughly one scale height. The evening spectrum displays strong features near 2.7~$\mu$m and 3.3~$\mu$m where the asymmetry is larger, the individual limb depths differ by over 3-$\sigma$, and the corresponding radii differ by up to two scale heights. In contrast, the morning limb spectrum appears relatively flat and lacks strong features. 

As discussed by ref.\cite{powell2019_LAmodels}, limb asymmetry can be more confidently attributed to an atmospheric signal, rather than systematics or ephemeris error, if larger asymmetry is observed at wavelengths corresponding to expected features in the transmission spectrum. The larger asymmetry we see around 2.7~$\mu$m matches an expected water absorption band, which was detected in WASP-107b's atmosphere by previous HST (6.5$\sigma$ significance) \cite{kreidberg2018_wasp107b} and JWST/MIRI observations ($>$12.5$\sigma$ significance) \cite{dyrek2023_wasp107b}. The presence of other molecules in WASP-107b's atmosphere is explored in ref.\cite{welbanks2024_wasp107b}. There is no reason to suggest water would not be present at both terminators, so the relatively flatter morning limb spectrum is likely the result of that limb being cooler which reduces the size of transmission features. As discussed further later, cloudiness likely plays a role too.

We ran several statistical tests to verify that the limb asymmetry we observe is real. First, we compared our transmission spectra to a ``null'' case of a model atmosphere with uniform-limbs using a $\chi^2$ test, which gave $\chi^2/\mathrm{dof}$ = 5.37. This rejects the uniform-limb model at high ($>$99\%) confidence. We also perform a paired sample t-test on the spectra, which returned a p-value of 4 $\times~10^{-11}$, and also strongly rejects our morning and evening spectra having identical means. Both tests assume the measured depths are uncorrelated across spectral channels, which we believe is true based on several further tests (Methods Section~\ref{subsec:methods_obsdetails}). We also verified that our asymmetric-limb light curve models fit the data better than a uniform-limb model with the same $t_c$ and total transit depth. Our asymmetric-limb models were always preferred by the reduced $\chi^2$, the Bayesian Information Criterion (BIC) by up to $\Delta$BIC = 10, and the Akaike Information Criterion (AIC) by up to $\Delta$AIC = 8.1. The Bayes Factor also preferred the asymmetric-limb model by factors of 1.1 - 137.6, corresponding to detection significances of 1.15 - 3.62$\sigma$ \cite{benneke2013_transmissionspectroscopy, welbanks2021_aurora}.

Next, we verified that our fits were not artificially injecting limb asymmetry. We generated thirty synthetic light curves of a uniform-limb planet with the same cadence, transit depth, and wavelength-dependent light curve scatter as our real observations. We drew the scatter from a Gaussian distribution, as our real data residuals are Gaussian based on an Anderson-Darling test. We fit these simulated observations with \texttt{catwoman} and repeated for ten realizations, extracting the limb spectra each time (Extended Data Figure~1). In almost all cases we recovered equal limb depths, and outliers with differing limb depths were random in wavelength and polarity, unlike our real observations. 

Finally, we checked for the signal of limb asymmetry directly in the data. Figure~\ref{fig:LC_and_residuals} shows data at 2.68~$\mu$m, where we see the largest limb asymmetry. We folded the light curve about the time of conjunction to compare the data in ingress and egress. If WASP-107b has asymmetric limbs, we would expect to see differences in the ingress and egress times caused by the larger or small sides of the planet reaching each contact point earlier or later than for a symmetric planet. Our observations indicate that the evening limb on WASP-107b is larger, so we would expect transit ingress and egress to occur late relative to a uniform-limb transit. In the folded data shown in Figure~\ref{fig:LC_and_residuals}, this would manifest as a positive difference between the observations taken during ingress and egress. We see this predicted signal directly in our data (Figure~\ref{fig:LC_and_residuals} Panel~A). Subtracting the data after transit center from that before transit center, we see a large and consistently positive difference during the ingress/egress period. 

The data also show significant residuals compared to a uniform-limb transit model during ingress and egress, as predicted by refs.\cite{line2016_LA, powell2019_LAmodels}. Panel~B of Figure~\ref{fig:LC_and_residuals} compares the residuals of our best-fit asymmetric-limb model to a uniform-limb model with the same transit depth. The residuals from the two fits match during mid-transit (between the second and third contact points) and the residuals are consistent with each other out-of-transit. However, the residuals of the asymmetric- and uniform-limb models do not match during ingress and egress, where we see the residuals to the uniform-limb model have a significantly non-zero mean of $198\pm61$\,ppm, while the mean of asymmetric-limb model's residuals is $30\pm61$\,ppm. This indicates that the uniform-limb model does not well fit the observations, and increases our confidence that that the observed ingress and egress of WASP-107b are indeed asymmetric.

\subsubsection{Ruling out alternate explanations} \label{subsec:alternatescenarios}

We can rule out several possible non-atmospheric origins for the limb asymmetry we infer. First, we investigated the impact of the measurement uncertainties in WASP-107b's orbital parameters and the star's limb-darkening coefficients. We repeated our light curve fits allowing these to freely vary in each channel, using Gaussian-shaped priors set by our joint-fit results and an ATLAS stellar model (Methods Section~\ref{subsec:fitting_LD}), as well as fully free limb-darkening coefficients. This did not change the resulting limb spectra and only increased the individual depth uncertainties by 1\% on average.


Next, we checked the impact of transit timing variations (TTVs). Previous observations have ruled out TTVs greater than 20 seconds \cite{mocnik2017_wasp107b}, but smaller TTVs must be considered given our sensitivity to timing precision. Using \texttt{TTVFaster} \cite{ttvfaster}, we estimate that the peak-to-peak TTV amplitude of WASP-107b due to its outer companion WASP-107c \cite{piaulet2021_wasp107b} is $<$0.4 seconds. This is smaller than our uncertainty on WASP-107b's $t_c$, and our spectra are robust to varying $t_c$ by this amount.


We also considered the timing accuracy of our instruments. Our strongest constraint on $t_c$ comes from the JWST observations. We consulted with JWST project scientists and verified that the image time-stamps (\texttt{INT\_TIMES}) are accurate within $<$0.5 seconds, and done consistently between instruments. We also verified that the onboard barycentric correction of these time-stamps fully accounts for the spacecraft position and velocity. For all JWST instruments, the \texttt{INT\_TIMES} values correspond to the time that the last pixel along the read direction is read-out. In our case, this occurs 1.347 s after the first pixel is read-out. In this work, we always used the mid-integration time, as given in the \texttt{INT\_TIMES} table, corrected to the middle of array read-out and converted to BJD TDB format. 

We also checked for excess correlated noise in our JWST/NIRCam data. The residuals between the data and the best-fit asymmetric-limb models were Gaussian distributed and passed Anderson-Darling and Shapiro-Wilks tests. From an Allan variance test, however, the root-mean-square of the residuals did deviate slightly from pure-white noise behavior, falling off as $N^{-0.4}$ with the number of integrations per bin ($N$) rather than $N^{-0.5}$. We also see minor fluctuations in the light curves of the detector reference pixels throughout the observation. Excess red noise may exist in our data, but we consider it very unlikely that red noise introduced a false limb asymmetry signal based on the injection-recovery tests we performed (Extended Data Figure~1).


Lastly, we consider the transit light source effect (TLSE). Stellar surface heterogeneities can affect the observed transmission spectrum \cite{rackham2018_tlseMstars, rackham2019_tlseFGKstars}, and spot crossings have been previously reported at optical wavelengths \cite{anderson2017_wasp107b, dai2017_wasp107b, mocnik2017_wasp107b, kreidberg2018_wasp107b} in transit observations of WASP-107b. However, we do not see evidence for a starspot crossing in our data. Based on TLSE models for a K6 star such as WASP-107\cite{rackham2019_tlseFGKstars}, unocculted starspots would change our measured transit depths by $\sim$50 ppm -- much smaller than the observed morning-to-evening differences. 

We also considered if a small starspot located right on the stellar limb could be causing the observed limb-asymmetry signal. If such a spot crossing is the sole source of the limb asymmetry, it would have to both offset the limb spectra and change their relative shapes. Due to limb-darkening, starspot crossings on the stellar limb have an order of magnitude weaker effect in the light curve than when the spot is occulted mid-transit, but may induce large ($\sim$10-100 s) TTVs \cite{oshagh2013_starspots}. If a limb-spot crossing were affecting our data, it would also bias the $t_c$ we derive from the simultaneous F210M observation (relative to the ancillary observations), but this is not the case (Methods Section~\ref{subsec:methods_tcfit}). 


\section{Discussion} \label{sec:discussion}

To date, limb asymmetry has been detected via ground-based high-resolution transmission spectroscopy on the T$_{eq} = $ 2150~K WASP-76b \citep{ehrenreich2020_wasp76b, kesseli2021_wasp76b, gandhi2022_wasp76b, maguire2024_wasp76b}, the 2260~K MASCARA-2b \citep{hoeijmakers2020_mascara2b}, and the 2720~K WASP-121b \citep{bourrier2020_wasp121b, borsa2021_wasp121b} via molecular abundance and wind speed gradients. WASP-107b joins the 1200~K WASP-39b [\cite{espinoza2024_wasp39b}, Delisle, S., et al. in review] as the only planets on which limb asymmetry has been detected via space-based low-resolution transmission spectroscopy, and separate limb transmission spectra recovered. Our discovery on WASP-107b is uniquely surprising because, at an equilibrium temperature of only 770 K, WASP-107b is significantly cooler than these other planets. Such cool atmospheres are not expected to exhibit strong spatial heterogeneity, as their circulation is expected to efficiently distribute heat globally, leading to generally low day-night contrast and homogeneous terminators \citep[e.g.][]{line2016_LA, kataria2016_gcmgrid, powell2019_LAmodels}. Our result suggests that either limb asymmetry is more predominant than current models predict, or WASP-107b is an outlier. 

The comprehensive modeling and the interpretation of our observational results is beyond the scope of this work, since it will likely require a unified treatment of 2D effects and clouds in WASP-107b's atmosphere. Nevertheless, we did compare our results to separate 1D atmosphere models for the morning and evening limbs to provide an initial estimate of the atmospheric parameters. 

We compared our observations to atmospheric models for the morning and evening terminator generated using \texttt{ScCHIMERA}. Described further in Methods Section~\ref{subsec:methods_forwardmodels}, we fixed each model to a metallicity of $\sim$10 times the Solar value and carbon-to-oxygen ratio of 0.35. These are the bestfit values from a combined retrieval analysis of previous HST/WFC3 and JWST/MIRI observations \cite{dyrek2023_wasp107b}, and are very consistent with the additional retrievals of ref.\cite{welbanks2024_wasp107b}. We include a vertically uniform gray ``cloud". The clouds or hazes in WASP-107b's atmosphere are likely more complex than this, but this simple treatment enables initially probing the relative role of clouds/hazes without adding more free parameters than is justified. We fit a grid of temperature and cloud opacity to the evening and morning limb spectra individually using the $\chi^2$-minimization method \cite{avni1976_chi2fit, practicalstatsforastrobook}. 

\begin{figure}[h!]
    \centering
    \includegraphics[width=\textwidth]{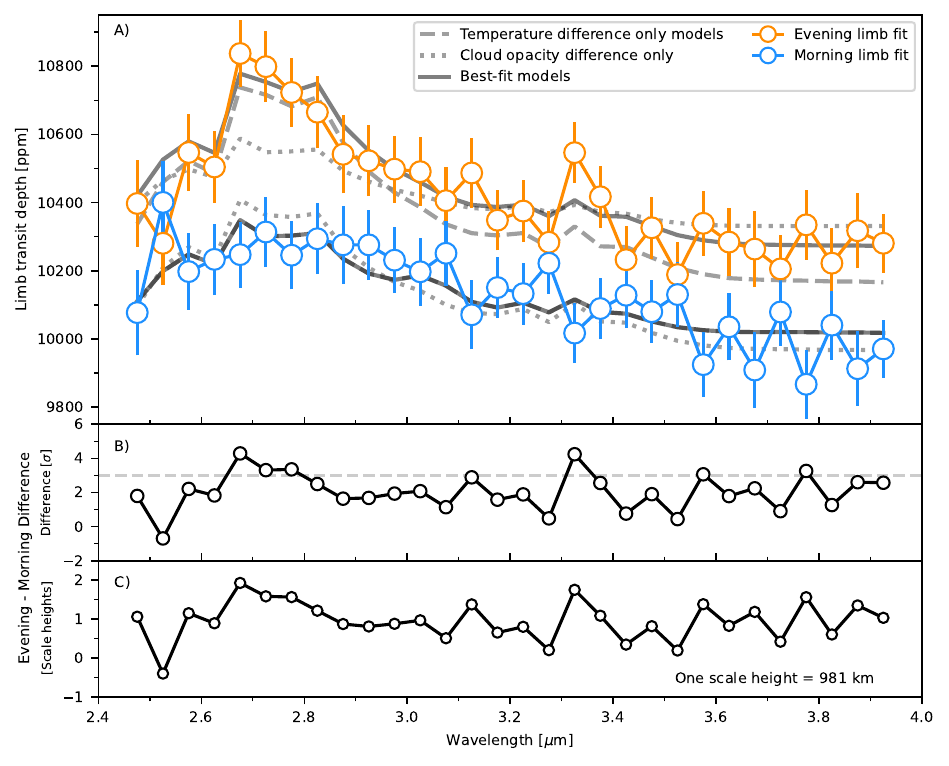}
    \caption{
    \textbf{The separate transmission spectra of WASP-107b's morning and evening limbs.} Panel A shows the transmission spectra of the evening (orange circles) and morning (blue circles) limbs, derived from our JWST/NIRCam F322W2 observations. Points with error bars are the median and standard deviations (1-$\sigma$, defined by the 16th and 84th percentiles) of the transit depth posterior distributions. We compare these to atmospheric models based on grid fits to these spectra (Figure~\ref{fig:gf_allreductions}). The solid gray lines show our best-fit solution where the evening limb is hotter by $\sim$180~K, and the limbs have different gray cloud opacity. To demonstrate the individual effects of temperature and cloud opacity, we show two additional model pairs that are not best-fit solutions, but exhibit the effect of changing just one parameter in an extreme way. These are the dashed lines, which share the same cloud opacity (10$^{-1.58}$~cm$^2$~g$^{-1}$) but differ in temperature by $\sim$190~K; and the dotted lines, which share the same temperature (713~K) but differ in log cloud opacity by 0.8~cm$^2$~g$^{-1}$. 
    The lower panels show the difference between the morning and evening depths in units of their mutual uncertainty (Panel B), and radii in units of WASP-107b's atmospheric scale height (Panel C). The limb depths are $>$1-$\sigma$ different at most wavelengths, and over 3-$\sigma$ different near 2.7~$\mu$m and 3.3~$\mu$m. WASP-107b's limb radii differ by 0.5 - 2 scale heights at these wavelengths. 
    }
    \label{fig:limbspectra}
\end{figure}

\begin{figure}[h!]
    \centering
    \includegraphics[width=\textwidth]{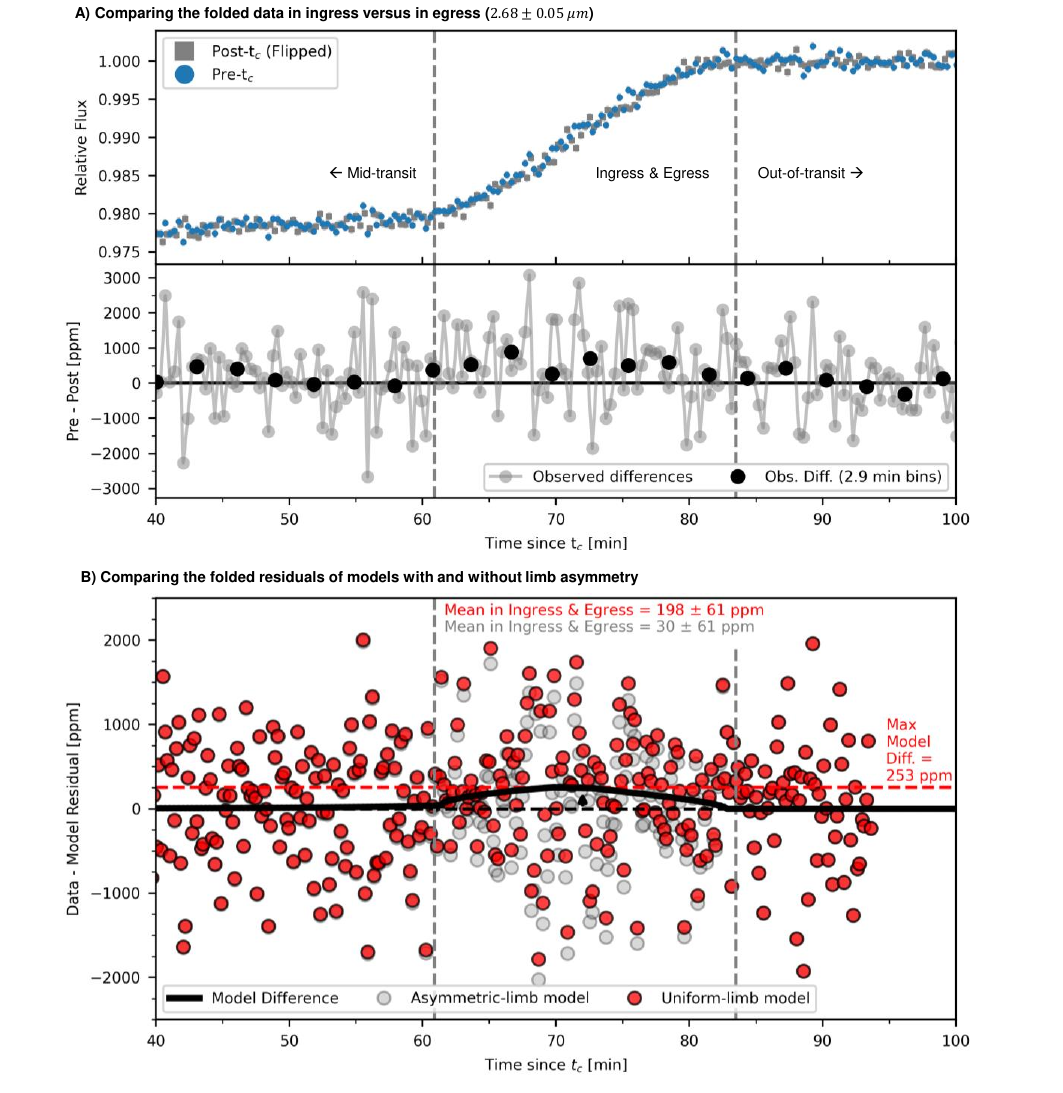}
    \caption{\textbf{Signal of limb asymmetry in our JWST/NIRCam data.} Here, we focus on our 2.68~$\mu$m channel which has the largest evening-morning difference (Figure~\ref{fig:limbspectra}). Panel~A shows the observed light curve data, with error bars represent the standard deviation (1-$\sigma$). Panel~B shows residuals from light curve models with and without limb asymmetry.  We fold each x-axis about the time of conjunction ($t_c$) so that ingress and egress are overlaid and more easily compared. We zoom in around the ingress/egress. In Panel~A, blue points show the data before $t_c$ while gray points show the data after $t_c$. We see a consistently positive difference between the pre- and post-$t_c$ points reaching $\sim$1000~ppm within ingress and egress, but no difference outside of ingress and egress, indicative of limb asymmetry. In Panel~B, gray points show the residuals of our data with the best-fit \texttt{catwoman} asymmetric-limb model at this wavelength (from which Figure~\ref{fig:limbspectra} is derived). The red points are residuals with a uniform-limb model, which reach a significantly larger mean value within ingress and egress. 
    }
    \label{fig:LC_and_residuals}
\end{figure}

\begin{figure}[h!]
    \centering
    \includegraphics[width=\textwidth]{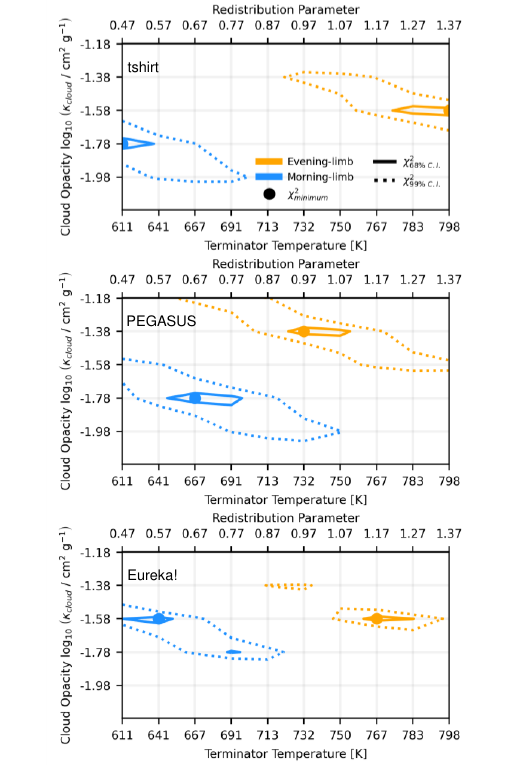}
    \caption{\textbf{Atmospheric model fits to WASP-107b's evening and morning limb spectra.}
    Shown are the results of our grid-fits described in Section~\ref{sec:discussion} and \ref{subsec:methods_forwardmodels} for the results of all three data reductions tried in this work. We primarily focused on the results from \texttt{tshirt} due to its more robust correction for 1/f noise, but the spread of parameter values between reductions is instructive for interpreting these values. We fixed the atmospheric metallicity to $\sim$10x the solar value and carbon-to-oxygen ratio to 0.35 based on the results of ref.\cite{dyrek2023_wasp107b} and ref.\cite{welbanks2024_wasp107b}, then fit for the redistribution parameter (i.e., terminator temperature, related as $T = 738.31 \times \text{redistribution}^{0.25}$) and gray cloud opacity for each limb individually. Results for the evening and morning limb are shown by the orange and blue contours, respectively, with contours drawn at the 68\% and 99\% confidence intervals. We find that the atmospheric conditions on each limb are significantly different. There is some correlation between the temperature and cloud opacity, but this difference in atmospheric conditions is robust. WASP-107b's morning limb is generally cooler than its evening limb, with a best-fit difference in temperature of 180 $\pm$ 36~K from \texttt{tshirt}, 68 $\pm$ 29~K from \texttt{PEGASUS}, and  137 $\pm$ 20~K from \texttt{Eureka!}. 
    }
    \label{fig:gf_allreductions}
\end{figure}

We find that WASP-107b's evening and morning terminators must have significantly different atmospheric properties, shown in Figure~\ref{fig:gf_allreductions}. There is some correlation between the effects of temperature and cloud opacity, but the evening and morning distributions do not overlap even far beyond their 99\% confidence intervals. The data prefer a cooler morning limb and a hotter evening limb, with a slight difference in the cloud opacity. The ``best-fit" models, corresponding to the minimum $\chi^2$ values, are marked by the points in Figure~\ref{fig:gf_allreductions} and compared to the data in Figure~\ref{fig:limbspectra}. Based on these best-fit values and 68\% confidence intervals for our \texttt{tshirt} reduction, we infer an evening-morning temperature difference of 180 $\pm$ 36~K. The precise difference is slightly reduction dependent with the \texttt{PEGASUS} and \texttt{Eureka!} spectra suggesting differences of 68 $\pm$ 29~K and 137 $\pm$ 20~K, respectively, but a hotter evening limb is always preferred. More spectroscopic data will help refine this measurement, explore the difference in cloudiness in more detail, and investigate whether limb asymmetry affects other molecules such as photochemically produced SO$_2$ \cite{dyrek2023_wasp107b, tsai2023_w39bSO2transport}.

The evening-morning temperature differences we derive are $\sim$10-23\% of WASP-107b's equilibrium temperature. Adapting a common metric for quantifying day-night heat redistribution on hot Jupiters (e.g.,\cite{perez2013_heatredistribution}), we compute A$_{e-m}$~$\approx$~0.33-0.64 as the proportional evening-morning energy difference, where A$_{e-m}$ $\equiv \left(F_{evening}-F_{morning}\right) / F_{evening}$ and $F_i = \sigma T_i^4$. Unfortunately, small sample size and differences in analyses make it difficult to compare WASP-107b in detail to the other planets on which limb asymmetry has been detected. WASP-76b is the only other planet with published evening-morning temperature differences; refs.\cite{gandhi2022_wasp76b, maguire2024_wasp76b} present large, varied estimates ranging 16 - 35\% of WASP-76b's equilibrium temperature, or A$_{e-m}$~$\approx$~0.38-0.83. Using the day- and nightside temperatures of WASP-121b from ref.\cite{bourrier2020_wasp121b} as proxies for its evening and morning temperatures, we can infer a difference of $>$25\% its equilibrium temperature, and A$_{e-m}$~$\approx$~0.65. Therefore, WASP-107b's limb asymmetry may be proportionally comparable to those of the much hotter WASP-76b and WASP-121b, though more data is needed before any population-level trends can be drawn. 

To illustrate the individual effects of temperature and cloudiness on the limb spectra, we show two additional model pairs in Figure~\ref{fig:limbspectra}. The dashed models share an opacity of 10$^{-1.58}$~cm$^2$~g$^{-1}$, midway between the distributions in Figure~\ref{fig:gf_allreductions}, and differ in temperature by $\sim$190~K (611~K and 798~K). The dotted models share a temperature of 713~K, again between the distributions, and differ in log cloud opacity by 0.8~cm$^2$~g$^{-1}$. These pairs roughly illustrate the limits of the distributions in Figure~\ref{fig:gf_allreductions} along each axis. Varying temperature has the strongest effect on the limb spectra, recreating both the offset and relative amplitude between morning and evening. Varying cloud opacity can recreate the offset but does not as well explain the difference in spectral amplitudes. In reality, both temperature and cloud/haze properties may change spatially and feedback on one another, highlighting the complex multidimensionality of exoplanet atmospheres. In this work, we demonstrate that unraveling this multidimensionality, and probing morning-to-evening differences is possible with JWST. This new capability opens an exciting new path toward a complete picture of these worlds beyond our own.

\clearpage

\clearpage

\section{Methods}\label{sec:methods}

\subsection{JWST/NIRCam Observation} \label{subsec:methods_obsdetails}

Our transit observation of WASP-107b using JWST/NIRCam was Observation 8 of program JWST-GTO-1185, which executed on January 14th, 2023. We acquired on our target WASP-107 using one 0.152 second integration with JWST/NIRCam imaging in its F335M filter, with \texttt{RAPID} readout on the \texttt{SUB32TATSGRISM} subarray. After target acquisition, we began our science observation consisting of simultaneous short-wavelength channel imaging with filter WLP8/F210M and long-wavelength channel spectroscopy with GRISMR/F322W2 \cite{NIRCamINSTRUMENT_Greene2010, NIRCam_inflightpaper}. We used the \texttt{BRIGHT2} readout pattern on subarray \texttt{SUBGRISM256}, and collected 1,293 integrations at 20.22 second cadence. Each integration consisted of seven groups. Our total exposure time was 26,145.53 seconds. 

Our reductions of these data using three different pipelines are described in the following section. From each reduction, we extracted thirty spectroscopic lightcurves from our F322W2 data in equally-spaced wavelength bins from 2.45 to 3.95~$\mu$m, corresponding to a spectral resolution of approximately 64. We chose to use this resolution because it offers a good trade-off between spectral information and transit depth measurement precision, as well as protects against optical correlations. At this low spectral resolution, we are binning together $\sim$28 detector resolution elements per channel. As a result, the point spread function is much smaller than our bin sizing and there is no optical correlation between channels. We further mitigate possible bin-to-bin correlations by correcting for 1/f noise, as described in the following Section.

\subsection{JWST/NIRCam Data Reduction} \label{subsec:methods_nircamdatareduction}

\subsubsection{Reduction with \texttt{tshirt}} \label{subsubsec:reduction_tshirt}

We used the \texttt{tshirt} pipeline (\url{https://tshirt.readthedocs.io}) to extract spectroscopic lightcurves of WASP-107b's transit.
As in ref.\cite{bell2023_w80}, we modified the CALDETECTOR1 stage of the \texttt{jwst} pipeline to turn the \texttt{\_uncal} files into rate images with less 1/$f$ noise.
We started with JWST pipeline version 1.8.4, CRDS Version 11.16.16, and CRDS context \texttt{jwst\_1039.pmap}.
\texttt{tshirt} replaces the default reference pixel correction with a row-by-row, odd/even by amplifier (ROEBA) correction \cite{schlawin2023SWNIRCamPerformance} using the background pixels, which can reduce the 1/$f$ noise as compared to reference pixels alone \cite{schlawin2020jwstNoiseFloorI}.
For the odd/even slow-read direction, we used the reference pixels in the bottom 4 rows.
For the row-by-row fast-read correction, we used all pixels from X=1846 to 2043 (0-based) to estimate the sky background and 1/f noise in that row.
For the jump step of the pipeline, we used a threshold of 6~$\sigma$.

We then continued the rest of the steps in CALDETECTOR1 with the default parameters.
After constructing rate files for each integration, we manually divided the images by a \texttt{jwst\_nircam\_flat\_0266.fits} imaging flat field and marked all pixels that have a ``DO\textunderscore NOT\textunderscore USE'' DQ value to NaN.
We multiplied all count rate images by the gain and integration time to estimate the total number of electrons at the end of the ramp for photon error calculation.
Finally, we cleaned cosmic rays from all images by combining 150 rate images at a time and marking all pixels that deviate from the median rate image by more than 20 times the \texttt{jwst} pipeline error for a given rate image for replacement.
The bad pixels are replaced by the linearly interpolated value of the 150-pixel time series.

We performed a column-by-column background subtraction with a linear fit along the Y direction to pixels Y=5 to 24 and Y=44 to 65 for all rate images.
For the spectral extraction, we first fit the spectrum row-by-row along the dispersion direction of integration 647 with a smooth 40-knot univariate spline with SciPy in order to build a profile function as a function of wavelength.
We used a co-variance weighted extraction \cite{schlawin2020jwstNoiseFloorI}, assuming a uniform correlation between pixels of 0.08 and a read noise of 14\,$e^-$.
We extracted a rectangular aperture from 29 to 39 pixels in the Y direction and 4 to 1747 in the X direction.
 While this rectangular aperture does not track the curvature of the trace and could influence the absolute flux and aperture corrections, we normalized the spectra by the out-of-transit baseline, so this effect is not expected to affect the derived planet spectra.
We used the profile along the Y direction to estimate the missing flux in pixels that marked as NaN and which deviate from the average profile by $\geq$30$\sigma$.

We fit the resulting lightcurves produced by \texttt{tshirt} in the same method as for all reductions, which is described in Section~\ref{subsec:fitting_asymlimb}. The best-fit transmission spectra are compared to those of our other reductions in Supplementary Figure~1.

\subsubsection{Reduction with \texttt{Eureka!}} \label{subsubsec:reduction_eureka}

Our reduction of the NIRCam F322W2 data with \texttt{Eureka!} (\url{https://eurekadocs.readthedocs.io/en/latest/}) followed the same procedure as ref.\cite{bell2023_w80}, which we summarize here. We used version 0.10.dev15+g80126b56.d20230613 of \texttt{Eureka!}\cite{EUREKAPIPELINE}, CRDS version 11.17.0 and context 1093, and \texttt{jwst} package version 1.10.2. The exact \texttt{Eureka!} Control Files and \texttt{Eureka!} Parameter Files we used are available for download (\url{https://zenodo.org/doi/10.5281/zenodo.12765422}).

\texttt{Eureka!}'s Stage 1 and 2 use the \texttt{jwst} pipeline for basic calibration. We ran both stages using their default settings, except for increasing the Stage 1 jump step's rejection threshold to 6.0 and skipping the photom step in Stage 2. In Stage 3, we cropped the frame to focus on relevant pixels (y-pixels 5--64 and x-pixels 15--1709), corrected for the curvature of the spectral trace, performed background subtraction per-column and per-integration, and then performed optimal spectral extraction using the pixels within 9 pixels of the flattened spectral trace. We then sigma-clipped (10$\sigma$ threshold) the observations compared to a 50-integration wide boxcar smoothed version of the observations to remove cosmic rays without removing sharp astrophysical features like the transit ingress/egress.

We fit the resulting lightcurves produced by \texttt{Eureka!} in the same method as for all reductions, which is described in Section~\ref{subsec:fitting_asymlimb}. The best-fit transmission spectra are compared to those of our other reductions in Supplementary Figure~1.

\subsubsection{Reduction with \texttt{Pegasus}} \label{subsubsec:reduction_TGB}

We also reduced and extracted the transmission spectrum of WASP-107b using the \texttt{Pegasus} pipeline (\url{https://github.com/TGBeatty/PegasusProject}). This pipeline is described in more detail in ref.\cite{beatty2024_gj3470b}, and we briefly summarize it here. We began with the \textsc{rateint} files from the \texttt{jwst} pipeline v1.10.2, using CRDS version 11.17.0 and context 1093. 

For each rateint file we first performed a background subtraction step by fitting a two-dimensional second-order spline to each integration using the entire 256$\times$2048 \textsc{rateint} images. We masked out image rows 5 to 75 to not self-subtract light from the WASP-107 system, and then we fit individual background splines to each of the four amplifier regions in the images. We then extrapolated the combined background spline over the masked portions near the star to perform the background subtraction. Visual inspection of the \textsc{rateint} images also showed that in roughly 5\% of the integrations the reference pixel correction failed for at least one of the amplifier regions, so after the spline fitting and subtraction we re-ran the reference pixel correction using \texttt{hxrg-ref-pixel} (\url{https://github.com/JarronL/hxrg_ref_pixels}). This appeared to correct the issue.

We then performed a spectrophotometric extraction on the background-subtracted images using optimal extraction techniques \cite{OptExtract}. We iteratively constructed a smoothed spatial profile for the F322W2 data using pixel columns from column 22 to 38 (inclusive) and pixel rows from row 5 to 1650 (inclusive), and we iteratively estimated a variance matrix for this same region starting from the pixel uncertainties and read-noise values following established techniques \cite{OptExtract}. We then fit a 4th-order polynomial spectral trace to each integration and extracted fluxes in each wavelength bin using an extraction aperture with a half-height of seven pixels centered on the estimated trace in each detector column. In doing so we accounted for partial pixel effects in both the spectral and spatial directions.

We fit the resulting lightcurves produced by \texttt{Pegasus} in the same method as for all reductions, which is described in Section~\ref{subsec:fitting_asymlimb}. The best-fit transmission spectra are compared to those of our other reductions in Supplementary Figure~1. 

\subsubsection{Comparing Reduction Pipelines} \label{subsubsec:reduction_comparison}

We reduced our JWST/NIRCam F322W2 data with three independent reduction pipelines described above, but ultimately present results from the \texttt{tshirt} reduction in our main analysis. Here, we compare the results from each reduction to show that while this choice is justified, it does not make a difference to our overall result. 

As in our primary analysis, we fit the spectroscopic light curves from each reduction with an asymmetric-limb transit model (see Section~\ref{subsec:fitting_asymlimb} for details) and derived morning and evening limb transmission spectra for each. All three reductions lead to the same qualitative result: an evening-limb spectrum that is generally at higher transit depth and displays stronger spectral features compared to the relatively flat morning-limb spectrum (Supplementary Figure~1). Comparing the transit depths in each individual channel, the morning and evening limb depths are all consistent within 1$\sigma$ between reductions in each individual channel, except for one outlier at $\sim$3.77~$\mu$m where \texttt{Eureka!} and \texttt{Pegasus} differ by $\sim$1.08-$\sigma$ (Supplementary Figure~1). There are no channels where \texttt{tshirt} disagrees with the other two reductions by more than 1-$\sigma$. The agreement between the evening limb spectra are particularly good, where the depths differ by less than 0.5 mutual standard deviations in most channels. The only notable difference between the three reductions' results is that the morning limb spectrum derived from \texttt{tshirt} is generally slightly lower than that of \texttt{PEGASUS} and \texttt{Eureka!} at short wavelengths, though never by more than 1-$\sigma$ as mentioned. A potential reason for this, as well as our justification of choosing \texttt{tshirt} as our primary focus, is described further below.

In the main text (Section~\ref{subsec:results_LA}), we presented several statistical tests to show that the morning and evening limb transmission spectra from the \texttt{tshirt} are significantly different. We repeated these tests for the other two reductions as well. Between the morning and evening spectra, we compute $\chi^2 / \mathrm{dof}$ values of 2.80 and 4.39 for the \texttt{PEGASUS} and \texttt{Eureka!} reductions, respectively. In both cases, for thirty degrees of freedom, the value of the survival function for the ``null" hypothesis that the morning and evening spectra are identical is effectively zero. Similarly, we compute paired t-test p-values of 2 $\times~10^{-6}$ and 1.3 $\times~10^{-7}$ for \texttt{PEGASUS} and \texttt{Eureka!}, respectively. In both cases, we again reject a case where the morning and evening limb spectra have identical means. We also repeated the grid-fit of model atmosphere spectra over temperature and gray cloud opacity, described in the Main text and Section~\ref{subsec:methods_forwardmodels}, for the \texttt{PEGASUS} and \texttt{Eureka!} spectra. The resulting $\chi^2$ distributions (e.g., as in Figure~\ref{fig:gf_allreductions}) overlapped, and tell a consistent story that the atmospheric properties on each limb are different. Regardless of which reduction is used, our result that WASP-107b's evening and morning limb transmission spectra are significantly different in the JWST/NIRCam F322W2 bandpass, and indicate a difference in temperature and cloudiness between terminators, holds true.

We ultimately chose to use \texttt{tshirt} in our primary analysis because, of the three pipelines, \texttt{t-shirt} performs the most robust correction for 1/f noise. This 1/f noise is one of, if not the most, dominant systematic noise source in NIRCam time series data using GRISMR readout \cite{schlawin2020jwstNoiseFloorI,schlawin2023SWNIRCamPerformance}. \texttt{Eureka!} and \texttt{PEGASUS} both adopt very similar techniques for 1/f noise correction, including a ``reference pixel step" where the mean value of each detector amplifier's reference (i.e., unexposed) pixels are subtracted from the value of the exposed pixels. This step is meant to eliminate overall bias offsets which may be different between amplifiers, as well as odd/even column effects, pre-amplifier resets, and 1/f noise. However, these reference pixels exhibit noise and do not sample frequencies greater than $\sim$200~Hz, reducing the effectiveness of this correction when used alone. Further, this step can occasionally fail, leaving systematic biases including 1/f noise present in the data. In these failure cases, \texttt{PEGASUS} then uses a custom routine \texttt{hxrg-ref-pixel} (linked in Section~\ref{subsubsec:reduction_TGB}) to re-correct the data, while \texttt{Eureka!} does no re-correction. On the other hand, \texttt{t-shirt} replaces this reference pixel step with row-by-row, odd/even by amplifier (ROEBA) correction, as described above and in ref.\cite{schlawin2023SWNIRCamPerformance}. ROEBA correction better corrects for and reduces 1/f noise in the data than the default correction \cite{schlawin2020jwstNoiseFloorI,schlawin2023SWNIRCamPerformance}. This difference in 1/f noise correction is most likely the reason for the slight differences between the spectra derived from \texttt{tshirt} and the other two pipelines, particularly at the shortest wavelengths of the morning limb spectra (Supplementary Figure~1). A similar slight offset in absolute transit depth between \texttt{tshirt} and \texttt{Eureka!} has been seen in other analyses, and was directly traceable to this 1/f noise correction \cite{bell2023_w80}. 

As a final comparison, we repeat the grid-fit of terminator temperature (via the redistribution parameter) and gray cloud opacity, as done for \texttt{tshirt} in the main text, on the spectra derived from \texttt{PEGASUS} and \texttt{Eureka!}. The results are shown in Figure~\ref{fig:gf_allreductions}. We find that, while the specific solution is slightly different between reductions, the general result remains the same. From all three reductions, we recover the result that WASP-107b's evening limb is hotter than its morning limb, and that there may be a difference in cloudiness or haziness as well. In all three cases, the evening and morning distributions do not overlap even well beyond their respective 99\% confidence intervals. Even marginalizing over the cloud opacity, the evening-morning temperature difference is significant at the respective 68\% confidence intervals for all three reductions. We find that \texttt{tshirt} gives the largest evening-morning temperature difference of 180 $\pm$ 36~K, \texttt{PEGASUS} gives the smallest difference of 68 $\pm$ 29~K, and \texttt{Eureka!} is in between with 137 $\pm$ 20~K. Of these, the difference derived by \texttt{tshirt} has the largest uncertainty. 

\subsection{Ancillary Data Reduction} \label{subsec:methods_ancillarydatareduction}

As mentioned in Section~\ref{sec:results}, we used additional observations of the WASP-107 system to support the analysis of our JWST/NIRCam observation. The radial velocity data we used included 31 measurements collected by ref.\cite{anderson2017_wasp107b} between 2011 and 2014 using the CORALIE spectrograph on the 1.2-meter Euler-Swiss telescope, in addition to 60 measurements collected by the California Planet Search (PI: A. Howard) for ref.\cite{piaulet2021_wasp107b} between 2017 and 2020 using the High Resolution Echelle Spectrometer \cite{KECK_HIRES} on the Keck I telescope. A full table of these radial velocity measurements, including uncertainties, is included in the electronic journal version of ref.\cite{piaulet2021_wasp107b}. We recorded these tabulated values, assuming all necessary data reduction and detrending was already performed, and used them in our analysis.  

TESS observed the WASP-107 system in its Sector 10, and captured three full transits and one partial transit of WASP-107b at a thirty-minute cadence between March and April 2019. We downloaded the publicly available light curves of these observations from Mikulski Archive for Space Telescopes, using the ``light curve" data files (file extension \texttt{.lc}) which have known systematics already removed from the data. The out-of-transit flux values exhibited no strong outstanding systematic trends. We discarded the partial transit and only used data around the three full transits. In order to improve computational efficiency, we only used data within $\pm$0.45 days of each transit midpoint which is sufficient out-of-transit baseline. 

We also observed one transit of WASP-107b using the Goodman High Throughput Spectrograph \citep{SOARGOODMANINSTRUMENT}, in imaging mode with an SDSS-i filter, on SOAR during program N23A-840705 (PI: M. Murphy). We reduced these observations using the AstroImageJ pipeline \citep{ASTROIMAGEJCODE}, which included comparison star detrending for basic telluric variation corrections. Despite this initial detrending, the light curve still displayed a significant, non-linear systematic trend so we first performed additional airmass detrending. The airmass of WASP-107 varied quadratically throughout our observation, reaching a minimum near the middle of our night. As a result, the flux of WASP-107 relative to the comparison stars was an asymmetric and non-injective function of airmass. That is, the relative flux at a given airmass took different values depending on whether the time was before or after the airmass minimum, likely the result of the telescope structure needing to flip orientation as well as the sky properties changing throughout the night. We fit two separate quadratic functions of the form
\begin{equation}
    f = p_2 x^2 + p_1 x + p_0
\end{equation}
to the data before and after the airmass minimum. Here, $f$ is the relative flux, $p_0$ - $p_2$ are fitting coefficients, $x \equiv z - z_{median}$ for the corresponding subset of data, and $z$ is the airmass. The best fit coefficients were $p_2$ = 0.2022, $p_1$ = -0.0374, and $p_0$ = 0.9958 for the pre-minimum set, and $p_2$ = -0.0066, $p_1$ = 0.01097, and $p_0$ = 1.0053 for the post-minimum set. This airmass detrending still did not remove all systematic trends from the light curve, so we applied a Gaussian Process (GP) using \texttt{George} \cite{GEORGEGPCODE} to fit the remaining residuals using an exponential-squared kernel. The best-fit amplitude and scale of this kernel function were 0.00860 and 0.00057, respectively. The results of these detrending fits are agnostic to the precise time of conjunction, so we were confident in performing this detrending prior to fully fitting this light curve simultaneously with the other data sets. The primary reason for doing so is computation efficiency, as the GP parameter sampling was too slow to justify including it in the full routine. 

Spitzer/IRAC has observed the WASP-107 system multiple times in both transit and eclipse. We downloaded the publicly available light curve of one transit observation of WASP-107b on April 26, 2017 using IRAC's channel 2 ($\sim$4.5~$\mu$m) observed for Program 13052 (PI: M. Werner) from the Spitzer Heritage Archive. This program also collected a transit observation using IRAC's channel 1 ($\sim$3.6~$\mu$m). We chose to only use the channel 2 dataset because its bandpass does not overlap with our F322W2 observations, both reducing duplicate information and providing more robustness against the bias of limb asymmetry on our time of conjunction measurement, and because reductions of channel 1 data are more susceptible to strong instrument systematics that would likely hinder getting a precise t$_c$ constraint from the data. Our reduction of the data generally follows the process of ref.\cite{beatty2018_spitzerreduction}. Starting with the basic calibrated data, we first perform background subtraction by masking the middle of each image, 10 pixels in from each edge, then subtracting the median value of the unmasked pixels. To precisely determine the position of the star in each image, we first performed three iterations of 5-$\sigma$ clipping on each pixel's time series to identify bad pixels, and replaced their values with the median value of their time series. Then, we used the Howell centroiding technique \cite{howell06_spitzerreduction} to determine the position of WASP-107 in each image and recorded it. Afterward, we returned to the pre-bad pixel correction images and performed photometric extraction. We extracted the flux within a 2.3-pixel radius circular aperture centered on WASP-107. We tested various aperture sizes and chose the one that minimized the resulting light curve scatter. The light curve showed a strong ramp at the beginning likely due to detector persistence, so we cut out the first 2,000 points. 

Finally, JWST/MIRI observed one transit of WASP-107b in Low Resolution Spectroscopy (LRS) mode during program JWST-GTO-1280 (PI: P.O.\ Lagage). We were granted use of the broad-band (5--12~$\mu$m) light curve of this transit observation for our analysis. The data was reduced using \texttt{Eureka!}, following the same procedure described in detail in the Eureka! v1 reduction section of ref.\cite{bell:inrevieww43}, which we briefly summarize here. We again use the \texttt{jwst} pipeline for Stage 1 and 2 with its default parameters, except for Stage 1's jump step rejection threshold which we increased to 8.0, and skipping Stage 2's photom step. In Stage 3, we extract pixels 11--61 along the spatial direction and 140--393 along the dispersion direction, not including any pixels marked as ``DO\_NOT\_USE" in the data quality array. We then perform background subtraction accounting for MIRI/LRS' ``cruciform artifact" by first removing bright scattered light rays via sigma-clipping, and excess periodic background noise and 1/$f$ noise by doing this subtraction per-column and per-integration. To further account for potential linear trends in the background flux along the detector, as in ref.\cite{bell:inrevieww43}, we used the mean from an equal number of pixels on either side of the spectral trace for each column and integration when computing the background level. Finally, we perform optimal extraction on the spectrum by calculating a median frame, clipping 5$\sigma$ outliers along the time axis, and smoothing the spectrum using a 7-pixel wide boxcar filter. In our reduction, we used a different custom linearity correction than ref.\cite{dyrek2023_wasp107b} developed for this same data, but we verified that this did not ultimately change our result.

\subsection{Lightcurve Fitting}
\label{subsec:methods_fitting}

\subsubsection{Limb Darkening Treatment} \label{subsec:fitting_LD}

Before fitting any of the data, we computed the expected limb darkening coefficients for each instrument's bandpass to use as priors. We used the online Exoplanet Characterization Toolkit Limb Darkening Calculator \cite{exoCTK} with the Kurucz ATLAS9 stellar model grid, with parameters for WASP-107 of T$_{eff}$ = 4425 K, log$\left(g\right)$ = 4.63 dex, and [Fe/H] = 0.02 dex. We computed the corresponding model quadratic limb darkening coefficients for each instrument's bandpass. The coefficient values pertaining to the ancillary data fit priors are listed in Extended Data Table~1, and all values including those used for the spectroscopic F322W2 light curves are listed in Supplementary Table~1. When fit freely with no connection between wavelengths, limb darkening coefficient profiles can take unphysical, non-smooth, non-self-consistent shapes as a function of wavelength. We use these model coefficients as Gaussian priors in our fit of the ancillary data to ensure that the coefficients are physical and self-consistent, while still allowing their uncertainty to be reflected in the ultimate result. We fix these coefficients when fitting the spectroscopic F322W2 data to reduce the number of free parameters, but verified that this decision did not affect our result (Section~\ref{subsec:alternatescenarios}).

\subsubsection{Uniform-limb transit model} \label{subsec:fitting_uniformlimb}

When fitting the data assuming or forcing the planet to have uniform limbs, we use the transit modelling package \texttt{batman} \citep{batman}. This code models the planetary disk in transit as a perfect circle, where the radius of this circle varies only with wavelength. We allowed the orbital parameters to be free for our ancillary data fit described below, and fixed them to the result of that fit (Extended Data Table~1) when analyzing our JWST/NIRCam F322W2 spectroscopic data.

\subsubsection{Ancillary Data Fits} \label{subsec:fitting_ancillary}

We fit the seven ancillary transit light curves simultaneously with the radial velocity data. The transit model we use for the transit observations is described in Section~\ref{subsec:fitting_uniformlimb}. Using the Markov Chain Monte Carlo method with the package \texttt{emcee}, we sampled the time of conjunction $t_c$, orbital period ($P$) as $\log_{10} \left( P \right)$, semi-major axis ($a$) as $\log_{10} \left( a / R_\star \right)$, orbital inclination ($i$) as $\cos \left( i \right)$, eccentricity ($e$) and argument of periastron ($\omega$) as $\sqrt{e} \sin \left( \omega \right)$ and $\sqrt{e} \cos \left( \omega \right)$, individual band uniform-limb planet-star radius ratios $R_p / R_\star$, quadratic limb darkening coefficients $u_1$ and $u_2$ for each instrument/filter, radial velocity semi-amplitude $K$, and radial system velocity $\gamma$. For each transit light curve except the SOAR data, we also applied a linear flux vs. time systematics ramp model of the form
\begin{equation}
    f = a_1 x + a_0,
\end{equation}
where $x \equiv t - t_{median}$, and we also fit for $a_1$ and $a_0$ of each individual light curve. We did not apply a linear systematic ramp to the SOAR data because any such systematic trend would have been removed from our pre-detrending of that data. 

When fitting data through a Bayesian inference method, such as MCMC, it is common practice to apply priors to fitting parameters based on previous, independent measurements. For our analysis, we must be careful to select priors which could not have been affected by unaccounted-for limb asymmetry. For this reason, we do not enforce any priors on the transit-related parameters except for the semi-major axis, inclination, and limb darkening coefficients. In other words, the values of these parameters do not factor into the prior probability value. Since the semi-major axis and inclination symmetrically affect the transit light curve on either side of $t_c$, these measurements should not be affected by unaccounted-for limb asymmetry. We therefore felt safe in applying priors to the semi-major axis and inclination, based on the independent measurements by ref.\cite{dai2017_wasp107b} using K2 observations. Similarly, limb darkening is a property of only the star and the instrument being used, so it would also not be affected. We applied Gaussian priors to each instrument's limb darkening coefficients based on an ATLAS stellar model, as described in Section~\ref{subsec:fitting_LD}. For the RV-related parameters, we did not apply any priors since the only other measurements (refs.\cite{anderson2017_wasp107b, piaulet2021_wasp107b}) are derived from the same data we reuse here.

We ran this sampling for 10,500 steps, which was sufficient for all parameters to converge. The resulting best-fit values and derived parameter values are listed in Extended Data Table~1. The data and best-fit models are plotted in Supplementary Figure~2, and we highlight our constraints on WASP-107b's orbital parameters in Supplementary Figure~3. 

\subsubsection{Asymmetric-limb model} \label{subsec:fitting_asymlimb}

When fitting the spectroscopic JWST/NIRCam F322W2 light curves and investigating for limb asymmetry, we used the transit modeling package \texttt{catwoman} \citep{Espinoza2021_catwoman2, catwoman1}, which models a planet with asymmetric limbs as the combination of two semi-circles with independent radii. This effectively splits the uniform disk modeled by \texttt{batman} in two along a central axis. The orientation of this axis with respect to the planet's orbital plane is an adjustable parameter in \texttt{catwoman}, and we fixed this axis to be aligned with the planet's orbital axis (see our discussion of this assumption in Section~\ref{subsec:methods_obliquity}).  
Again, we fixed the orbital parameters and limb darkening coefficients, and sampled the time of conjunction, two planet-star radius ratios per wavelength channel, as well as the slope and intercept of a linear flux vs. time ramp. As mentioned in the Main Text, we applied a Gaussian prior to the time of conjunction based on the posterior distribution from our ancillary data fit. We ran each channel's MCMC sampling for 10,000 steps, which was more than sufficient for all parameters to converge. 

\subsubsection{WASP-107b's Obliquity and Orbital Misalignment} \label{subsec:methods_obliquity}

Fits of the current radial velocity measurements of WASP-107 favor WASP-107b having a small but non-zero eccentricity \cite[][and this work]{piaulet2021_wasp107b}. Further, starspot crossing frequencies \cite{dai2017_wasp107b} and measurements of the Rossiter-McLaughlin effect in this system \cite{rubenzahl2021_wasp107b} suggest that WASP-107b may be on a misaligned orbit. This eccentricity and misalignment are indicative of dynamical interactions that occurred in the planet's past, and may still be operating today. Here, we discuss the impact of these effects on our experimental design and results.

From our fit of the radial velocity data, we measure WASP-107b's eccentricity to be e = 0.05~$\pm$ 0.01 (Extended Data Table~1), consistent with that of ref.\cite{piaulet2021_wasp107b} which analyzed the same data. Neither this eccentricity, or any parameter in our fit, is affected by the Rossiter-McLaughlin (RM) effect. The RM effect only affects radial velocity measurements made during the planet's transit, and all of the measurements we use were made outside of WASP-107b's transit. Also, the measurement of the RM effect in this system by ref.\cite{rubenzahl2021_wasp107b} was done by assuming and fixing WASP-107b's orbital parameters to those derived by ref.\cite{piaulet2021_wasp107b}, with whom we are fully consistent. Relatedly, the detailed models of ref.\cite{carteret2023_rmeffectandtspectroscopy} show that the RM effect does not bias transmission spectra observed by JWST, especially at the very low spectral resolution that we bin our data to.

In deriving our limb spectra (Figure~\ref{fig:limbspectra}), our \texttt{catwoman}-based transit models \cite{catwoman1, Espinoza2021_catwoman2} assume that WASP-107b's rotation axis is aligned to its orbital axis. Tidal interaction models find that both rotational axis alignment (i.e., zero rotational obliquity) and synchronization (i.e., equal rotation and orbital periods) should be inevitable and happens very quickly for short period (P $<$ 10~days) gaseous planets \cite{hut1981_tidalevolution, peale1999_obliquity, fabrycky2007_obliquity, correia2010_tidalevolution}. Further, this alignment and synchronization can occur even if the orbit has not fully circularized \cite{peale1999_obliquity}. First derived by ref.\cite{guillot1996_obliquity} and rewritten by ref.\cite{rauscher2023_warmjupiters}, the timescale $\tau$ for this alignment can be estimated as 
\begin{equation}
    \tau \approx 0.067 \times \left(\frac{Q}{10^5}\right)\left(\frac{\omega_p M_p R_J^3}{\omega_J M_J R_p^3}\right) \left( \frac{P}{10~day}\right)^4~ Gyr.
\end{equation}
Here, $Q$ is the tidal quality factor, and $\omega$, $M$, and $R$ are the rotation rate, mass, and radius of the planet (subscript $p$) and Jupiter (subscript $J$). Assuming WASP-107b is synchronously rotating, so that $\omega_p = 2.02 \times 10^{-6}~s^{-1}$, we calculate an alignment timescale of $\sim$980~years for a Neptune-like $Q$ of $10^4$, or $\sim$9800~years for a Jupiter-like $Q$ of $10^5$. This is extremely short in either case, especially compared to its orbital circularization timescale of $\sim$66~Myr \cite{piaulet2021_wasp107b} and the estimated 3.4 - 8.3~Gyr age of the system \cite{mocnik2017_wasp107b, piaulet2021_wasp107b}. Therefore, even if WASP-107b's orbit is misaligned to its star's rotation axis, the planetary rotation axis should almost certainly be aligned.

\subsection{Constraining WASP-107b's time of conjunction} \label{subsec:methods_tcfit}

As described in Section~\ref{subsec:fitting_ancillary}, we fit the ancillary transit and radial velocity observations described in Sections~\ref{subsec:results_observations} and \ref{subsec:methods_ancillarydatareduction} together simultaneously. The goal of this fit was to precisely constrain WASP-107b's orbit before fitting our NIRCam/F322W2 spectroscopic data. These orbital parameters must be known precisely a priori to enable accurately separating the contribution of each limb to the observed transmission signal. Uncertainty in the time of conjunction ($t_c$), in particular, is degenerate with the effect of limb asymmetry on the planet's transit light curve \cite{line2016_LA, vonparis2016_LA, powell2019_LAmodels, Espinoza2021_catwoman2}. Using these ancillary data, we were able to strongly constrain each of WASP-107b's main orbital parameters (Extended Data Table~1). We highlight the resulting posterior distributions of several primary orbital parameters from this fit in Supplementary Figure~3. Our best-fit values are consistent with previous measurements from various other datasets \cite{dai2017_wasp107b, mocnik2017_wasp107b, anderson2017_wasp107b, piaulet2021_wasp107b, kokori2022exoclock} and, most importantly, we achieve very high precision on WASP-107b's $t_c$. At the epoch of our NIRCam F322W2 observation, we measure WASP-107b's $t_c$ to a 1-$\sigma$ precision of 0.70 seconds. This tight constraint is enabled by the wide wavelength coverage of these ancillary data, and the unprecedented combination of high photometric precision and fast integration cadence of the included JWST data.

The degeneracy between limb asymmetry and $t_c$ offsets naturally raises the question whether our best-fit value may be biased by limb asymmetry in our ancillary data. Our approach combining multi-wavelength measurements ensures such bias is eliminated, even if present in individual channels. This multi-wavelength approach is the current best method based on discussions of this problem in the literature \cite{powell2019_LAmodels, Espinoza2021_catwoman2}, and has been shown to eliminate any limb-asymmetry induced bias to the measured $t_c$ even when each data-set individually exhibits limb asymmetry of the same polarity \cite{powell2019_LAmodels}. Ref.\cite{powell2019_LAmodels} describes the effectiveness of this approach very well, which we summarize here. This method takes advantage of the fact that limb asymmetry, even if present at all wavelengths, will be a chromatic effect -- the asymmetry will be different at different wavelengths. When fit simultaneously, each wavelength will attempt to ``pull" $t_c$ by different amounts which, since $t_c$ is shared between every wavelength, leads to general disagreement between channels and a poor fit overall. In effect, the offset in each channel works against each other. The best solution is the true geometric time of conjunction, which was proven to be recovered in each test of this method by ref.\cite{powell2019_LAmodels}.

This approach to deriving $t_c$ assumes that limb asymmetry varies chromatically, though this cannot be empirically known a priori. However, the fact that we derive different morning and evening spectra in one data set (Figure~\ref{fig:limbspectra}) is proof that the exact same limb asymmetry is not uniformly present at all wavelengths. For instance, when fitting our broadband NIRCam F322W2 data by itself assuming a uniform-limb model and allowing $t_c$ to freely vary, we get a best-fit $t_c$ that is 6.64 $\pm$ 1.91~seconds later than the joint-fit value (Supplementary Figure~4). If the exact same limb asymmetry were present at all wavelengths and the biases did not cancel out as previously described, then our joint-fit would have given a time offset by this same value. If we then applied this offset time in our \texttt{catwoman} fits to the F322W2 spectroscopic data, we would not have observed any significant limb asymmetry. In this case, the assumed time of conjunction would be biased by an amount corresponding to the underlying asymmetry, which would have erased the asymmetry's signal.

To further evaluate our joint-fit's effectiveness, we tested fitting each ancillary data set individually using a uniform-limb model and allowing $t_c$ to freely vary. We then compared the difference between each individually fit $t_c$ and the joint-fit $t_c$. As just mentioned, when fitting the broadband JWST/NIRCam F322W2 data, we get a best-fit $t_c$ that 6.64 $\pm$ 1.91~seconds later than the joint-fit result from the broadband light curve, and we also see chromatic offsets in the best-fit $t_c$ shown in Supplementary Figure~4 from the spectroscopic data. Both the SOAR and Spitzer/IRAC Ch.2 observations had differences consistent with zero (7.6 $\pm$ 19.7~s and 4.42 $\pm$ 7.39~s, respectively). The TESS data, which overlaps with the SOAR bandpass, had a best-fit offset of -33.5 $\pm$ 19.0~s. This is negative and non-zero at over 1$\sigma$ significance, but it is not clear whether such large offset is real given the sparse 30-minute cadence of the data. The JWST/NIRCam F210M data, which was observed simultaneously with the NIRCam F322W2 data with the same cadence, had an offset of -1.11 $\pm$ 1.06~s. This is also negative but nearly consistent with zero. Finally, the JWST/MIRI LRS data had a difference of 2.62 $\pm$ 1.28~s which is non-zero by just over 2$\sigma$. Due to their superior data quality, our joint-fit is most effected by the JWST/NIRCam F210M and JWST/MIRI LRS observations. These individual fits show that both data sets may also be subject to some limb asymmetry, though not to the same degree as the JWST/NIRCam F322W2 data, and potentially of opposite polarity. Our multi-wavelength approach will therefore work as described to mitigate these individual biases. All together, these fits underscore the importance of combining data at multiple wavelengths. The mean offset of these individual fit values, weighted by their uncertainties ($1 / \sigma_{t_c}^2$), is 0.41~s which is well within the joint-fit value's uncertainty. Even if we ignore the largest offset, which was the TESS value, the mean offset is still only 0.47~s. To the best of our current knowledge, the $t_c$ we derive from our joint-fit is as accurate and bias-free as possible.

The above discussion raises an additional question of whether there is some systematic error in our JWST/NIRCam F322W2 data that causes the derived transit time to be different from the joint-fit value, and thus be the cause of the limb asymmetry we observe. We can directly compare the F322W2 data to the F210M data as these were observed simultaneously. Both data sets were reduced using the same method, so there is no difference in how the data time-stamps are calculated or corrected (e.g., barycentric correction). NIRCam's short (F210M) and long (F322W2) channels use separate detectors and the observatory software does not command synchronization, leading to slight differences in the exact timestamps between each light curve. On average, the timestamps of our F322W2 points are 0.41~s later than the corresponding F210M points. This timestamp offset is within the uncertainty of our joint-fit derived $t_c$, so it cannot explain the 6.64 $\pm$ 1.91~s $t_c$ difference derived from the F322W2 data. To verify this, we tested adding this timestamp offset to the F210M light curve points, then repeating the individual uniform-limb fit. The resulting offset was not significantly different (now -0.76 $\pm$ 1.08~s). We got a similar result when instead testing subtracting this offset from the F322W2 points. Therefore, there is no evidence that any such systematic errors are present and driving our result.  

\subsection{Forward Modeled Transmission Spectra} \label{subsec:methods_forwardmodels}

We compared our observational results to 1D atmosphere models for the morning and evening limbs to provide an initial estimate of the atmospheric parameters on each terminator. The model transmission spectra presented in Figure~\ref{fig:limbspectra} and discussed in the main text were generated using the \texttt{ScCHIMERA} model. This model was first described in ref.\cite{piskorz2018_schimera} with additional updates in refs.\cite{gharibnezhad2019_schimera, mansfield2021_schimera, iyerline2023_schimera}, but we provide a short summary here. See also ref.\cite{bell2023_w80} for a more detailed description and most recent updates (including opacity sources). In short, the \texttt{ScCHIMERA} tool is a self-consistent 1D radiative-convective-photochemical equilibrium solver (1D-RCPE). Chemistry can be computed in either thermochemical equilibrium or coupled with the VULCAN chemical kinetics code \cite{tsai2017_vulcan} to self-consistently compute disequilibrium abundances arising from photochemistry and vertical mixing. We initially produced a grid of 4830 WASP-107 specific chemical-disequilibrium 1D-RCPE models over a range of metallicities (23 points from -0.5 - +2.25, where the value is a base-10 logarithm), redistributions (scaling to the incident stellar flux, 14 points from 0.46 - 1.4 where 1 is full, 2 is dayside), and carbon-to-oxygen ratios (15 points from 0.1 - 0.75).  To fit the spectra we post-process the atmospheric structure (T-P profile and gas volume mixing ratios) with the \texttt{CHIMERA} \cite{line2013_chimera} transmission spectrum tool while also adjusting for an opaque gray, vertically uniform cloud. We parameterize this cloud as a ``gray" cross-section that is added to the gas absorption cross-sections. It can be taken as the product of a cloud droplet mixing ratio times an absorption cross-section but does not consider cloud microphysics.

From the grid of models described above, we used an atmospheric metallicity of 10 times the solar value and carbon-to-oxygen ratio of 0.35 based on the best-fit retrieval values on HST/WFC3 and JWST/MIRI observations of this planet from ref.\cite{dyrek2023_wasp107b}. These values are also consistent with the additional retrievals in ref.\cite{welbanks2024_wasp107b}. We expect the metallicity and carbon-to-oxygen ratio to be globally uniform quantities, and do not vary them between the models for each limb. In this work, we do not seek to make specific, precise constraints on global properties, but rather develop an understanding for what may be causing the observed difference between the evening and morning limbs. To do this, we performed a grid fit over the heat redistribution parameter, over its full range, and the gray cloud opacity, ranging from 10$^{-2.58}$ to 10$^{-0.78}$~cm$^2$~g$^{-1}$. This heat redistribution parameter is effectively the temperature at the terminator, related by $T = 738.31 \times \text{redistribution}^{0.25}$. 
We computed the $\chi^2$ value between the grid model and the observed morning and evening spectra individually, following the $\chi^2$-minimization technique of refs.\cite{avni1976_chi2fit, practicalstatsforastrobook}.

The results of our grid fit are shown in Figure~\ref{fig:gf_allreductions} and discussed in the main text. The best-fit model spectra are shown along with the observed limb spectra in Figure~\ref{fig:limbspectra}. In Figure~\ref{fig:gf_allreductions}, we show the $\chi^2$ distributions of the morning-limb fit in blue and the evening-limb fit in orange via contours drawn at the 68\% (i.e., 1-$\sigma$) and 99\% (i.e., 3-$\sigma$) confidence intervals. The distributions for each limb do not overlap even well beyond their 99\% confidence intervals, showing that the atmospheric properties on WASP-107b's terminators are significantly different. The morning-limb distribution is against the lower edge of our temperature grid with a minimum $\chi^2$ at 611~K (redistribution parameter of 0.47), and is below 630~K (667~K) at 68\% (99\%) confidence. The morning-limb uniform gray cloud opacity is well constrained between 10$^{-1.98}$ and 10$^{-1.58}$~cm$^2$~g$^{-1}$ at 99\% confidence, though as we have noted this is a simple initial exploration and the true nature of clouds or hazes on WASP-107b are likely more complex, and beyond the scope of this work. The evening-limb distribution is bimodel, with a cooler/clearer solution (T $\approx$ 732~K, $\kappa_{cloud} \approx 10^{-0.78}$~cm$^2$~g$^{-1}$) and a hotter/cloudier solution (T $\approx$ 760~K, $\kappa_{cloud} \approx 10^{-1.58}$~cm$^2$~g$^{-1}$). In either case, these fits suggest that the evening-limb is significantly hotter than the morning-limb, and that there is a slight difference in the cloudiness as well. 

Highlighted by the apparent bimodality of our evening-limb distribution, we see a correlation between the temperature and cloud opacity. This makes sense as both generally affect the amplitude of spectral features. To get a better sense for the individual effect of each parameter, we show two additional pairs of models in Figure~\ref{fig:limbspectra} -- one with fixed temperature but varied cloud opacity, and vice-versa. As discussed in the Main text, we find that the temperature has the strongest effect on both the relative shapes of the evening and morning spectra, as well as their relative offset. In our models, the cloud opacity (at least within this range of interest) also affects the offset, but does not as strongly change the shape of each spectrum. This precise behavior is obviously subject to our parametrization of these clouds as a vertically uniform gray opacity source. For example, more detailed cloud or haze models, especially focused on specific species, may affect the shapes differently or even induce relative slopes across the spectra. Disentangling these details will require additional observations over a wider wavelength coverage as well as much more complex multi-dimensional modeling. Nevertheless, our results show that recovering the separate properties of a transiting exoplanet's morning and evening terminators is possible, and that WASP-107b exhibits significant morning-to-evening asymmetry. 

\backmatter

\begin{appendices}
\section{Extended Data}\label{secA1}


\begin{sidewaystable}
\caption{\textbf{Priors for fitting the ancillary radial velocity and transit observations of WASP-107b, as well as the best-fit parameter values from these fits.} The fitted time of conjunction corresponds to the actual epoch of our JWST/NIRCam F210M / F322W2 transit observation. Note that we use the convention for $\omega$ used by the \texttt{RadVel} code \cite{RADVELpaper}. } 
\centering
\label{tab:ancillary_results}
\begin{tabular}{lcccl}\toprule 
Parameter & Units & Value  \\ \midrule 
Priors --- & & \\
$t_c$ & Time of Conjunction [BJD$_{\text{TDB}}$] & None  \\
P & Orbital Period [day] &  None    \\
 a / R$_\star$ & Semi-major axis & 18.164 $\pm$ 0.037 \cite{dai2017_wasp107b} \\
i  & Inclination [deg] & 89.8 $\pm$ 0.2 \cite{dai2017_wasp107b}   \\
e & Eccentricity & None &   \\
$\omega$ & Argument of Periastron [deg] & None &  \\
$\left( u_1, u_2 \right)_{\text{TESS}}$ & Quadratic LD coefficients, TESS &  $\left( 0.541 \pm 0.005, 0.106 \pm 0.007\right)$  \\
$\left( u_1, u_2 \right)_{\text{i-band}}$ & Quadratic LD coefficients, i-band &  $\left( 0.535 \pm 0.001, 0.120 \pm 0.001\right)$  \\
$\left( u_1, u_2 \right)_{\text{F210M}}$ & Quadratic LD coefficients, NIRCam F210M & $\left( 0.13 \pm 0.01, 0.26 \pm 0.01\right)$  \\
$\left( u_1, u_2 \right)_{\text{IRAC}}$ & Quadratic LD coefficients, IRAC Ch.2 & $\left( 0.10 \pm 0.01, 0.12 \pm 0.01\right)$  \\
$\left( u_1, u_2 \right)_{\text{MIRI}}$ & Quadratic LD coefficients, MIRI broad-band &  $\left( 0.076 \pm 0.001, 0.089 \pm 0.001\right)$ 
\\ \\
Astrophysical Parameter Best-fit Values --- & &   \\
$t_c$  & Time of Conjunction [BJD$_{\text{TDB}}$] & 2459958.747244 $\pm$ 8.2 $\times$ 10$^{-6}$  \\
$\log_{10} \left(P\right)$ & log Orbital period [days] & 0.75750893 $\pm$ 2.9 $\times$ 10$^{-8}$ \\
$\log_{10} \left(a / R_\star \right)$ & log semi-major axis [R$_\star$] & 1.25639 $\pm$ 0.0024 \\
$\cos ~i$  & Cosine of orbital inclination &  0.0074 $\pm$ 0.0005   \\
$R_p / R_\star$ & Radius ratio, TESS band &  0.1421 $\pm$ 0.0005 \\
$R_p / R_\star$ & Radius ratio, i-band (SOAR/Goodman) & 0.1435 $\pm$ 0.0003  \\
$R_p / R_\star$ & Radius ratio, JWST/NIRCam F210M &  0.143982 $\pm$ 0.00006 \\
$R_p / R_\star$ & Radius ratio, Spitzer/IRAC Ch.2 & 0.1407 $\pm$ 0.0004  \\
$R_p / R_\star$ & Radius ratio, JWST/MIRI broadband & 0.14368 $\pm$ 0.00004 \\
$K$ & RV semi-amplitude [m/s] &  13.74 $\pm$ 0.31 \\
$\gamma$ & System velocity [km/s] & -0.0022 $\pm$  0.0002  \\
$\sqrt{e} \cos \left( \omega \right) $ & & 0.234 $\pm$ 0.024   \\
$\sqrt{e} \sin \left( \omega \right) $ & & -0.009 $\pm$ 0.024 \\
& & & \\
Derived Parameter Values --- & & \\
$P$ & Orbital period [days] & 5.72148722 $\pm$ 3 $\times$ $10^{-7}$ \\
$a / R_\star$ & Semi-major axis [R$_\star$] & 18.05 $\pm$ 0.1  \\
$i$ & Orbital inclination [deg] & 89.57 $\pm$ 0.03  \\
$e$ & Eccentricity & 0.05 $\pm$ 0.01 \\
$\omega$ & Argument of Periastron [deg] & -2.3 $\pm$ 6.1 
  \\\botrule 
\end{tabular}
\end{sidewaystable}

\begin{sidewaystable}
    \centering 
    \caption{\textbf{Best-fit limb transit depths from our fit to WASP-107b's JWST/NIRCam F322W2 transit light curves}. The combined limb depth is the sum of the evening and morning limb depths, accounting for their covariance. Note that we used constant wavelength spacing for our bins.}
    \label{tab:bestfitdepths}
    \begin{tabular}{lcccccl}
    \toprule 
    Wavelength [$\mu$m] & Evening limb depth [ppm] & Morning limb depth [ppm] & Total transit depth [ppm]  & Limb depths covariance\\\midrule 
    2.475 $\pm$ 0.025 & 10397 $\pm$ 126 & 10077 $\pm$ 125 & 20474 $\pm$ 49 & -1.5 $\times$ 10$^{-8}$  \\ 
    2.525 $\pm$ 0.025 & 10280 $\pm$ 122 & 10400 $\pm$ 121 & 20681 $\pm$ 39 & -1.4 $\times$ 10$^{-8}$  \\ 
    2.575 $\pm$ 0.025 & 10546 $\pm$ 111 & 10197 $\pm$ 111 & 20744 $\pm$ 43 & -1.2 $\times$ 10$^{-8}$  \\ 
    2.625 $\pm$ 0.025 & 10504 $\pm$ 105 & 10233 $\pm$ 105 & 20737 $\pm$ 45 & -1.0 $\times$ 10$^{-8}$  \\ 
    2.675 $\pm$ 0.025 & 10837 $\pm$ 97 & 10247 $\pm$ 96 & 21084 $\pm$ 40 & -8.6 $\times$ 10$^{-9}$  \\ 
    2.725 $\pm$ 0.025 & 10798 $\pm$ 104 & 10313 $\pm$ 103 & 21112 $\pm$ 39 & -1.0 $\times$ 10$^{-8}$  \\ 
    2.775 $\pm$ 0.025 & 10722 $\pm$ 101 & 10245 $\pm$ 99 & 20968 $\pm$ 42 & -9.2 $\times$ 10$^{-9}$  \\ 
    2.825 $\pm$ 0.025 & 10665 $\pm$ 106 & 10293 $\pm$ 104 & 20959 $\pm$ 42 & -1.0 $\times$ 10$^{-8}$  \\ 
    2.875 $\pm$ 0.025 & 10541 $\pm$ 114 & 10276 $\pm$ 115 & 20817 $\pm$ 47 & -1.2 $\times$ 10$^{-8}$  \\ 
    2.925 $\pm$ 0.025 & 10522 $\pm$ 103 & 10275 $\pm$ 103 & 20798 $\pm$ 37 & -1.0 $\times$ 10$^{-8}$  \\ 
    2.975 $\pm$ 0.025 & 10498 $\pm$ 98 & 10231 $\pm$ 96 & 20729 $\pm$ 34 & -8.9 $\times$ 10$^{-9}$  \\ 
    3.025 $\pm$ 0.025 & 10490 $\pm$ 100 & 10197 $\pm$ 99 & 20687 $\pm$ 36 & -9.4 $\times$ 10$^{-9}$  \\ 
    3.075 $\pm$ 0.025 & 10405 $\pm$ 97 & 10251 $\pm$ 94 & 20656 $\pm$ 41 & -8.4 $\times$ 10$^{-9}$  \\ 
    3.125 $\pm$ 0.025 & 10486 $\pm$ 102 & 10070 $\pm$ 101 & 20557 $\pm$ 32 & -9.9 $\times$ 10$^{-9}$  \\ 
    3.175 $\pm$ 0.025 & 10348 $\pm$ 88 & 10150 $\pm$ 88 & 20499 $\pm$ 36 & -7.2 $\times$ 10$^{-9}$  \\ 
    3.225 $\pm$ 0.025 & 10375 $\pm$ 90 & 10133 $\pm$ 90 & 20508 $\pm$ 35 & -7.6 $\times$ 10$^{-9}$  \\ 
    3.275 $\pm$ 0.025 & 10283 $\pm$ 91 & 10222 $\pm$ 89 & 20505 $\pm$ 34 & -7.6 $\times$ 10$^{-9}$  \\ 
    3.325 $\pm$ 0.025 & 10546 $\pm$ 88 & 10017 $\pm$ 87 & 20563 $\pm$ 34 & -7.2 $\times$ 10$^{-9}$  \\ 
    3.375 $\pm$ 0.025 & 10417 $\pm$ 91 & 10089 $\pm$ 90 & 20506 $\pm$ 35 & -7.6 $\times$ 10$^{-9}$  \\ 
    3.425 $\pm$ 0.025 & 10232 $\pm$ 97 & 10128 $\pm$ 96 & 20360 $\pm$ 40 & -8.6 $\times$ 10$^{-9}$  \\ 
    3.475 $\pm$ 0.025 & 10326 $\pm$ 91 & 10079 $\pm$ 91 & 20406 $\pm$ 38 & -7.7 $\times$ 10$^{-9}$  \\ 
    3.525 $\pm$ 0.025 & 10188 $\pm$ 94 & 10130 $\pm$ 94 & 20319 $\pm$ 38 & -8.2 $\times$ 10$^{-9}$  \\ 
    3.575 $\pm$ 0.025 & 10339 $\pm$ 95 & 9924 $\pm$ 95 & 20264 $\pm$ 34 & -8.6 $\times$ 10$^{-9}$  \\ 
    3.625 $\pm$ 0.025 & 10284 $\pm$ 98 & 10035 $\pm$ 98 & 20319 $\pm$ 43 & -8.7 $\times$ 10$^{-9}$  \\ 
    3.675 $\pm$ 0.025 & 10263 $\pm$ 111 & 9908 $\pm$ 112 & 20172 $\pm$ 48 & -1.1 $\times$ 10$^{-8}$  \\ 
    3.725 $\pm$ 0.025 & 10205 $\pm$ 99 & 10079 $\pm$ 98 & 20284 $\pm$ 40 & -9.0 $\times$ 10$^{-9}$  \\ 
    3.775 $\pm$ 0.025 & 10334 $\pm$ 101 & 9867 $\pm$ 100 & 20201 $\pm$ 38 & -9.5 $\times$ 10$^{-9}$  \\ 
    3.825 $\pm$ 0.025 & 10222 $\pm$ 102 & 10040 $\pm$ 100 & 20262 $\pm$ 44 & -9.4 $\times$ 10$^{-9}$  \\ 
    3.875 $\pm$ 0.025 & 10317 $\pm$ 110 & 9912 $\pm$ 109 & 20230 $\pm$ 34 & -1.1 $\times$ 10$^{-8}$  \\ 
    3.925 $\pm$ 0.025 & 10280 $\pm$ 85 & 9970 $\pm$ 84 & 20251 $\pm$ 35 & -6.6 $\times$ 10$^{-9}$  \\ 
         \botrule 
    \end{tabular}
\end{sidewaystable}

\begin{figure}[ht!]
    \centering
    \includegraphics[width=\textwidth]{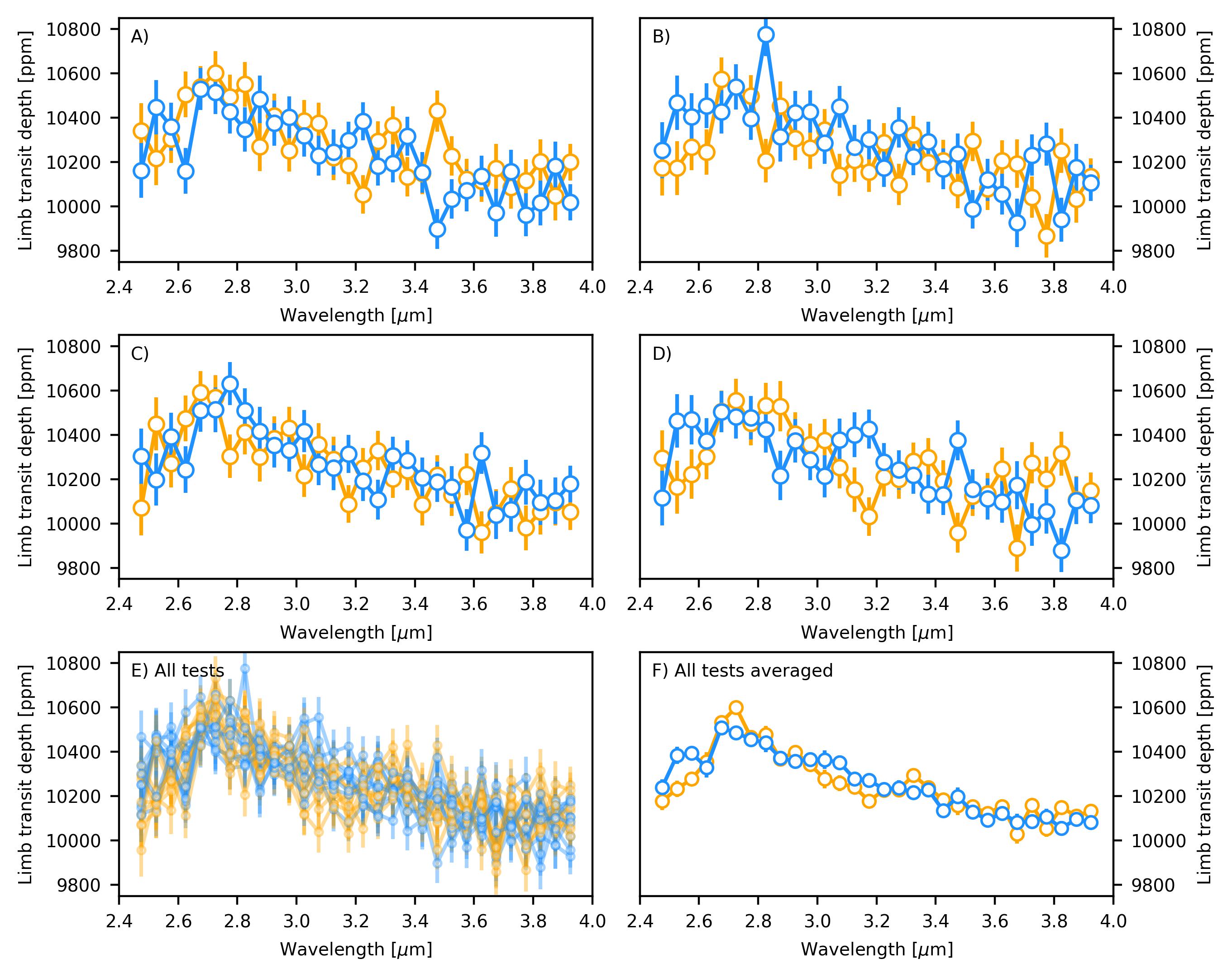}
    \caption{
    \textbf{Uniform-limb injection test results.}
    To better interpret our observed limb spectra (Figure~\ref{fig:limbspectra}), we tested what we would derive if there were no limb asymmetry at all. As described in Section~\ref{subsec:results_LA}, we generated several realizations of synthetic data with the same Gaussian light curve scatter and cadence of our real observations, based around a transit model with identical morning and evening limbs. Then, we fit these synthetic data with \texttt{catwoman} as we did with the real data. As in our presentation of the real data in Figure~\ref{fig:limbspectra}, we show the best-fit evening-limb spectrum in orange and the best-fit morning-limb spectrum in blue.
    The top four panels show the results of four individual tests. The bottom left panel shows all ten tests overlaid on top of each other, and the bottom right panel shows the average limb depth in each channel. In almost all cases, we recover the absence of limb asymmetry in each channel and the best-fit evening and morning transit depths are consistent with one another within 1-$\sigma$. Due to noise in the data, there are occasional outliers where the depths differ by more than 1-$\sigma$, but these occur randomly and have no consistent or regular pattern with wavelength. Also, these random variations switch polarity at random, further suggesting they cannot be real. 
    }
    \label{fig:injectiontests}
\end{figure}
\clearpage
\section{Supplementary Information}\label{secB1}
\begin{figure}[ht!]
    \centering
    \includegraphics[width=\textwidth]{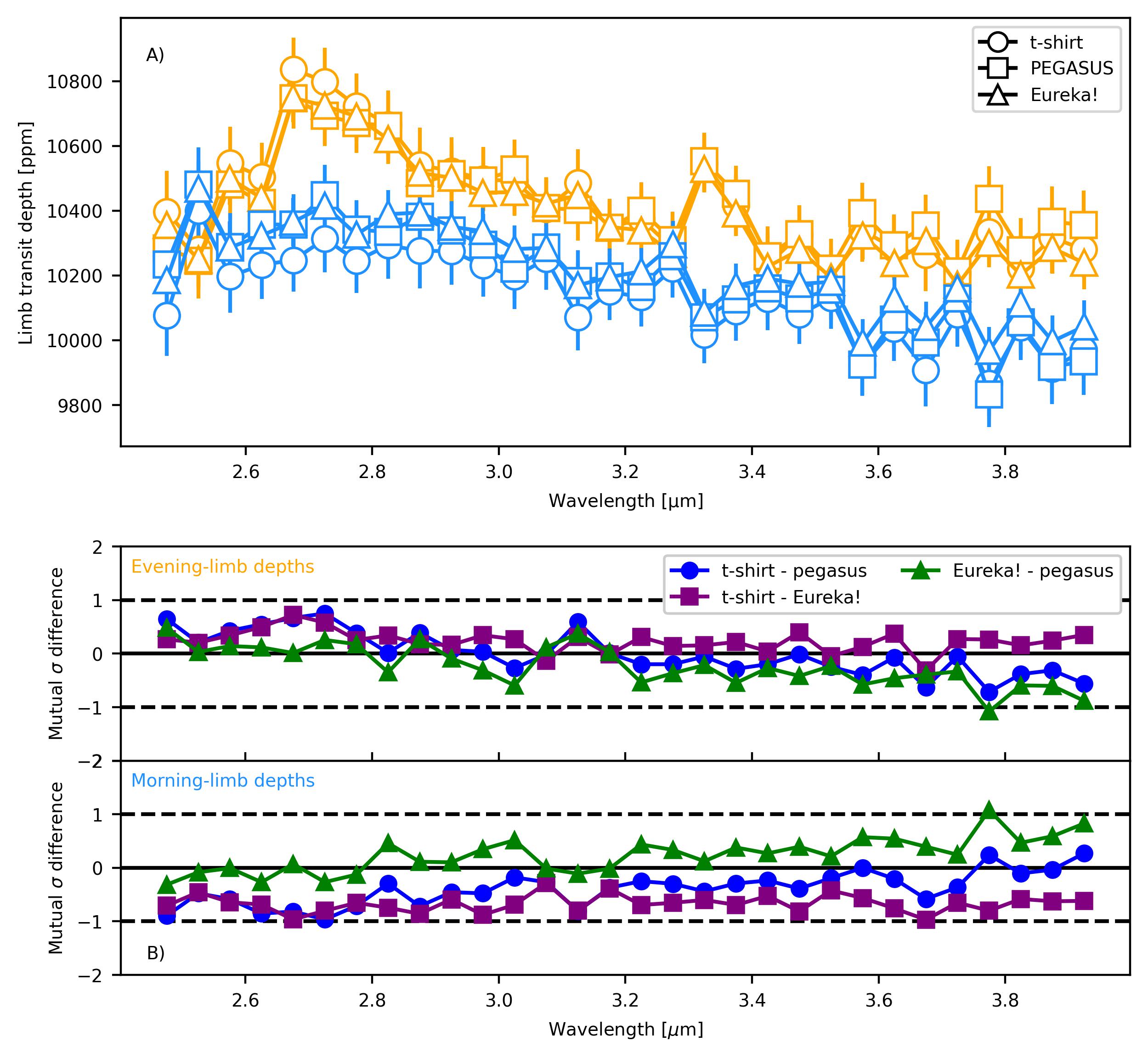}
    \caption{
    \textbf{Comparison of the transmission spectra determined for WASP-107b's morning and evening limbs from each of our reduction methods.} 
    \textit{Top:} the morning and evening limb transmission spectra derived from each reduction of our JWST/NIRCam F322W2 observation. The results from \texttt{t-shirt} are shown as circles, \texttt{Pegasus} as squares, and \texttt{Eureka!} as triangles. Each reduction produces morphologically consistent spectra for both limbs. 
    \textit{Bottom:} the difference, in terms of the mutual standard deviation, between the limb transit depths derived by each reduction. Those between the evening-limb depths are shown in the upper panel, and those between the morning-limb depths in the lower panel. The exact limb transit depths in each channel vary slightly between reductions, but are all consistent within 1-$\sigma$ except for one outlier. At shorter wavelengths, the morning-limb spectrum from \texttt{t-shirt} is slightly lower than that of the other two reductions due to \texttt{t-shirt}'s improved 1/f noise correction, though this difference is always less than 1-$\sigma$. In our main analysis, we ultimately chose to use the \texttt{t-shirt} reduction due to this improved 1/f noise correction. 
    }
    \label{fig:comaringreductions}
\end{figure}

\begin{table}[ht!]
    \centering 
    \caption{\textbf{Quadratic limb darkening coefficients used in this analysis}. We computed all coefficients using the Kurucz ATLAS9 stellar model grid available through the online Exoplanet Characterization Toolkit. }
    \label{tab:LDCs}
    \begin{tabular}{lcccccl}
    \toprule 
    Instrument & Wavelength(s) [$\mu$m] & u$_1$ & u$_2$ & Prior type \\\midrule 
        TESS & 0.78 $\pm$ 0.21 & 0.541 $\pm$ 0.005 & 0.106 $\pm$ 0.007 & Gaussian  \\
        SOAR/Goodman & 0.77 $\pm$ 0.10 & 0.535 $\pm$ 0.001 & 0.120 $\pm$ 0.001 & Gaussian \\
        JWST/NIRCam F210M & 2.09 $\pm$ 0.20 & 0.13 $\pm$ 0.01 & 0.26 $\pm$ 0.01 & Gaussian \\
        JWST/NIRCam F322W2 & 2.475 $\pm$ 0.025 & 0.140 $\pm$ 0.012 & 0.221 $\pm$ 0.017 & Fixed  \\ 
                & 2.525 $\pm$ 0.025 & 0.141 $\pm$ 0.012 & 0.215 $\pm$ 0.016 & Fixed  \\ 
                & 2.575 $\pm$ 0.025 & 0.136 $\pm$ 0.012 & 0.214 $\pm$ 0.016 & Fixed  \\ 
                & 2.625 $\pm$ 0.025 & 0.133 $\pm$ 0.012 & 0.211 $\pm$ 0.016 & Fixed  \\ 
                & 2.675 $\pm$ 0.025 & 0.130 $\pm$ 0.012 & 0.209 $\pm$ 0.016 & Fixed  \\ 
                & 2.725 $\pm$ 0.025 & 0.127 $\pm$ 0.012 & 0.208 $\pm$ 0.016 & Fixed  \\ 
                & 2.775 $\pm$ 0.025 & 0.125 $\pm$ 0.012 & 0.205 $\pm$ 0.016 & Fixed  \\ 
                & 2.825 $\pm$ 0.025 & 0.123 $\pm$ 0.012 & 0.202 $\pm$ 0.016 & Fixed  \\ 
                & 2.875 $\pm$ 0.025 & 0.122 $\pm$ 0.012 & 0.198 $\pm$ 0.016 & Fixed  \\ 
                & 2.925 $\pm$ 0.025 & 0.122 $\pm$ 0.011 & 0.195 $\pm$ 0.015 & Fixed  \\ 
                & 2.975 $\pm$ 0.025 & 0.121 $\pm$ 0.011 & 0.192 $\pm$ 0.015 & Fixed  \\ 
                & 3.025 $\pm$ 0.025 & 0.120 $\pm$ 0.011 & 0.189 $\pm$ 0.015 & Fixed  \\ 
                & 3.075 $\pm$ 0.025 & 0.119 $\pm$ 0.011 & 0.186 $\pm$ 0.015 & Fixed  \\ 
                & 3.125 $\pm$ 0.025 & 0.118 $\pm$ 0.011 & 0.183 $\pm$ 0.015 & Fixed  \\ 
                & 3.175 $\pm$ 0.025 & 0.118 $\pm$ 0.011 & 0.180 $\pm$ 0.014 & Fixed  \\ 
                & 3.225 $\pm$ 0.025 & 0.118 $\pm$ 0.011 & 0.177 $\pm$ 0.014 & Fixed  \\ 
                & 3.275 $\pm$ 0.025 & 0.115 $\pm$ 0.010 & 0.176 $\pm$ 0.014 & Fixed  \\ 
                & 3.325 $\pm$ 0.025 & 0.115 $\pm$ 0.010 & 0.173 $\pm$ 0.014 & Fixed  \\ 
                & 3.375 $\pm$ 0.025 & 0.115 $\pm$ 0.010 & 0.170 $\pm$ 0.014 & Fixed  \\ 
                & 3.425 $\pm$ 0.025 & 0.114 $\pm$ 0.010 & 0.167 $\pm$ 0.013 & Fixed  \\ 
                & 3.475 $\pm$ 0.025 & 0.114 $\pm$ 0.010 & 0.165 $\pm$ 0.013 & Fixed  \\ 
                & 3.525 $\pm$ 0.025 & 0.114 $\pm$ 0.010 & 0.163 $\pm$ 0.013 & Fixed  \\ 
                & 3.575 $\pm$ 0.025 & 0.113 $\pm$ 0.010 & 0.161 $\pm$ 0.013 & Fixed  \\ 
                & 3.625 $\pm$ 0.025 & 0.112 $\pm$ 0.010 & 0.159 $\pm$ 0.013 & Fixed  \\ 
                & 3.675 $\pm$ 0.025 & 0.110 $\pm$ 0.009 & 0.157 $\pm$ 0.013 & Fixed  \\ 
                & 3.725 $\pm$ 0.025 & 0.109 $\pm$ 0.009 & 0.155 $\pm$ 0.012 & Fixed  \\ 
                & 3.775 $\pm$ 0.025 & 0.109 $\pm$ 0.009 & 0.153 $\pm$ 0.012 & Fixed  \\ 
                & 3.825 $\pm$ 0.025 & 0.108 $\pm$ 0.009 & 0.151 $\pm$ 0.012 & Fixed  \\ 
                & 3.875 $\pm$ 0.025 & 0.108 $\pm$ 0.009 & 0.150 $\pm$ 0.012 & Fixed  \\ 
                & 3.925 $\pm$ 0.025 & 0.107 $\pm$ 0.009 & 0.148 $\pm$ 0.012 & Fixed  \\
        Spitzer/IRAC & 4.49 $\pm$ 0.50 & 0.10 $\pm$ 0.01 & 0.12 $\pm$ 0.01 & Gaussian \\
        JWST/MIRI  & 5-- 12 & 0.076 $\pm$ 0.001 & 0.089 $\pm$ 0.001 & Gaussian \\
         & \\
         \botrule 
    \end{tabular}
\end{table}

\begin{figure}[ht!]
    \centering
    \includegraphics[width=\textwidth]{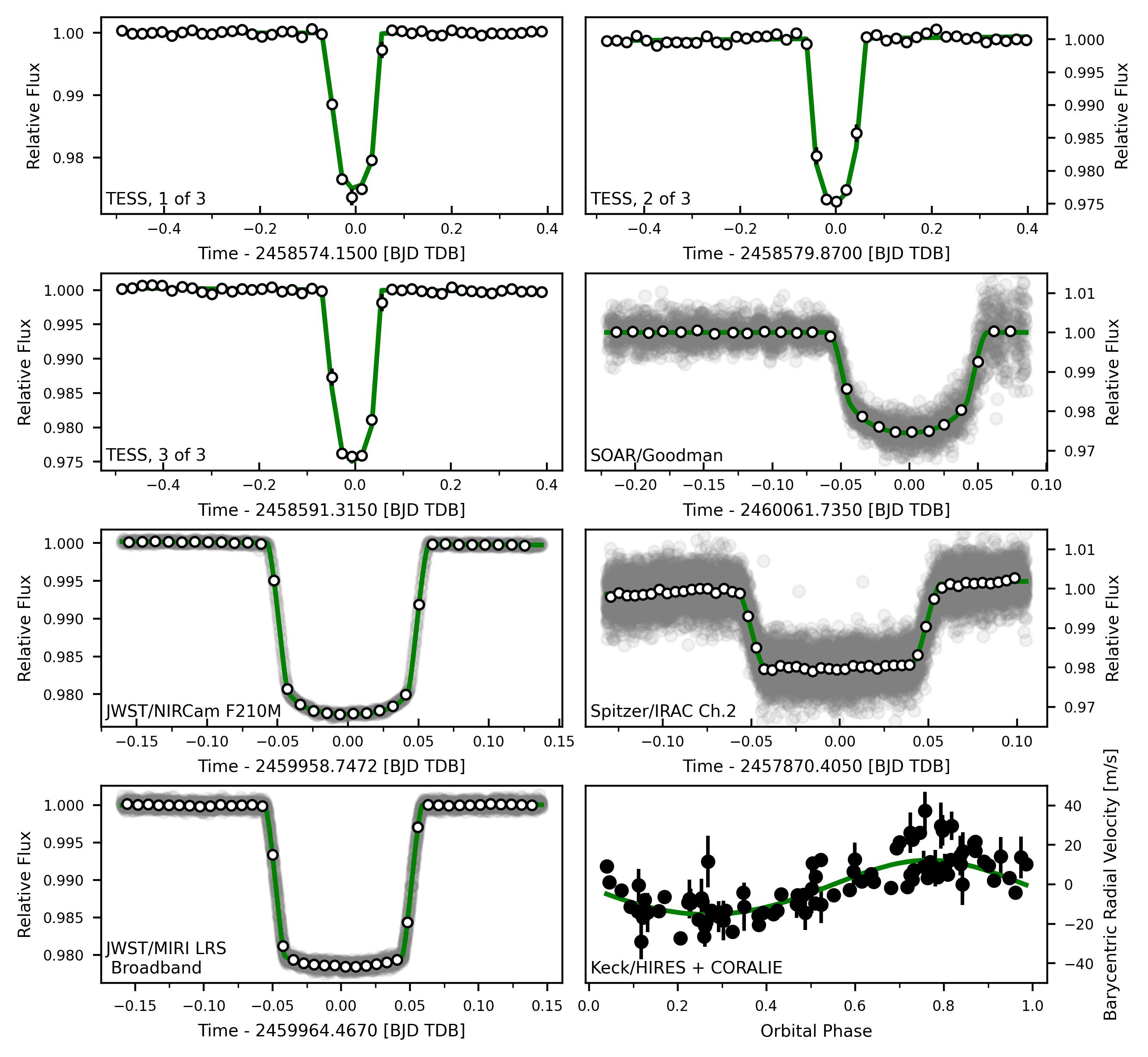}
    \caption{\textbf{Gallery of ancillary transit and radial velocity data used to support our analysis.} We collected archival transit observations from TESS and Spitzer/IRAC, observed new data with SOAR/Goodman and JWST/NIRCam F210M, and received broadband JWST/MIRI LRS data from JWST program 1280 from P.I. Pierre-Olivier Lagage. The data for TESS are shown as the white points. The data for the other transit observations are shown as the gray points, with temporally binned data as the white points. The best-fit models, including the transit model and systematics models combined, are shown in green. We also collected archival radial velocity measurements observed from Keck/HIRES and CORALIE \cite{anderson2017_wasp107b, piaulet2021_wasp107b}. These data are shown phase-folded in the bottom-right panel, and the best-fit model is shown in green. We used these transit and radial velocity data to constrain WASP-107b's orbital parameters for use in fitting our JWST/NIRCam F322W2 spectroscopic light curves. In particular, with these data we constrained WASP-107b's time of conjunction, at the epoch of our NIRCam observation, to a 1-$\sigma$ precision of 0.7 seconds. We see no significant evidence for starspot crossings during any of our transit observations.}
    \label{fig:ancillary_gallery}
\end{figure}

\begin{figure}[ht!]
    \centering
    \includegraphics[width=\textwidth]{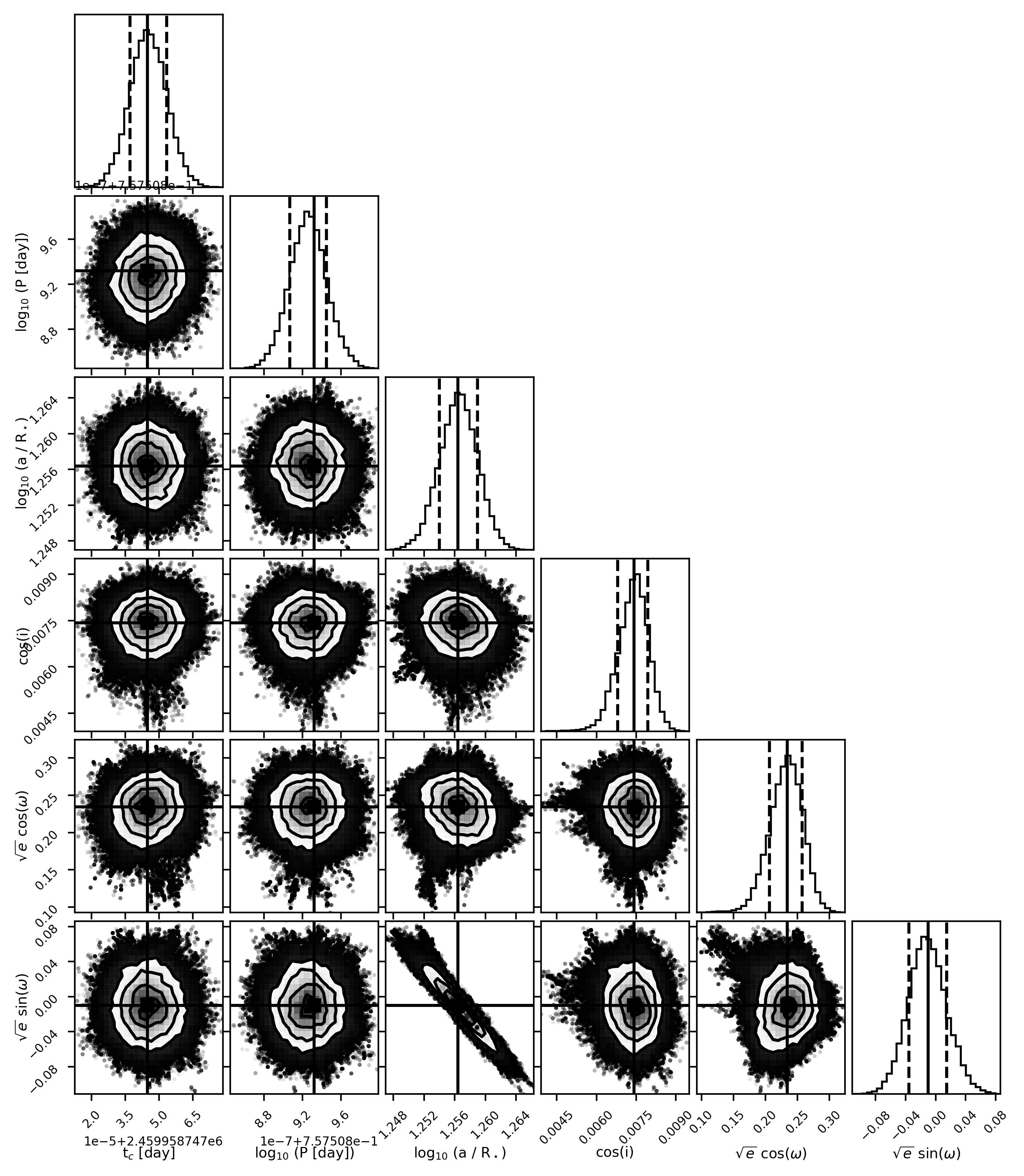}
    \caption{
    \textbf{Posterior distributions of WASP-107b's primary orbital parameters from our ancillary data joint-fit.} As described in the Methods, we simultaneously fit transit observations from TESS, SOAR/Goodman, JWST/NIRCam F210M, Spitzer/IRAC, and JWST/MIRI LRS, as well as radial velocity observations from Keck/HIRES and CORALIE \cite{anderson2017_wasp107b, piaulet2021_wasp107b}. These data and best-fit models are shown in Figure~\ref{fig:ancillary_gallery}, and here we show the posterior distributions for the main planetary orbital parameters. Each parameter was well constrained. The primary goal of this fit was to measure the time of conjunction precisely, shown in the left-most panel, which we did to a 1-$\sigma$ precision of only 0.70~seconds.}
    \label{fig:orbital_cornerplot}
\end{figure}

\begin{figure}
    \centering
    \includegraphics[width=\textwidth]{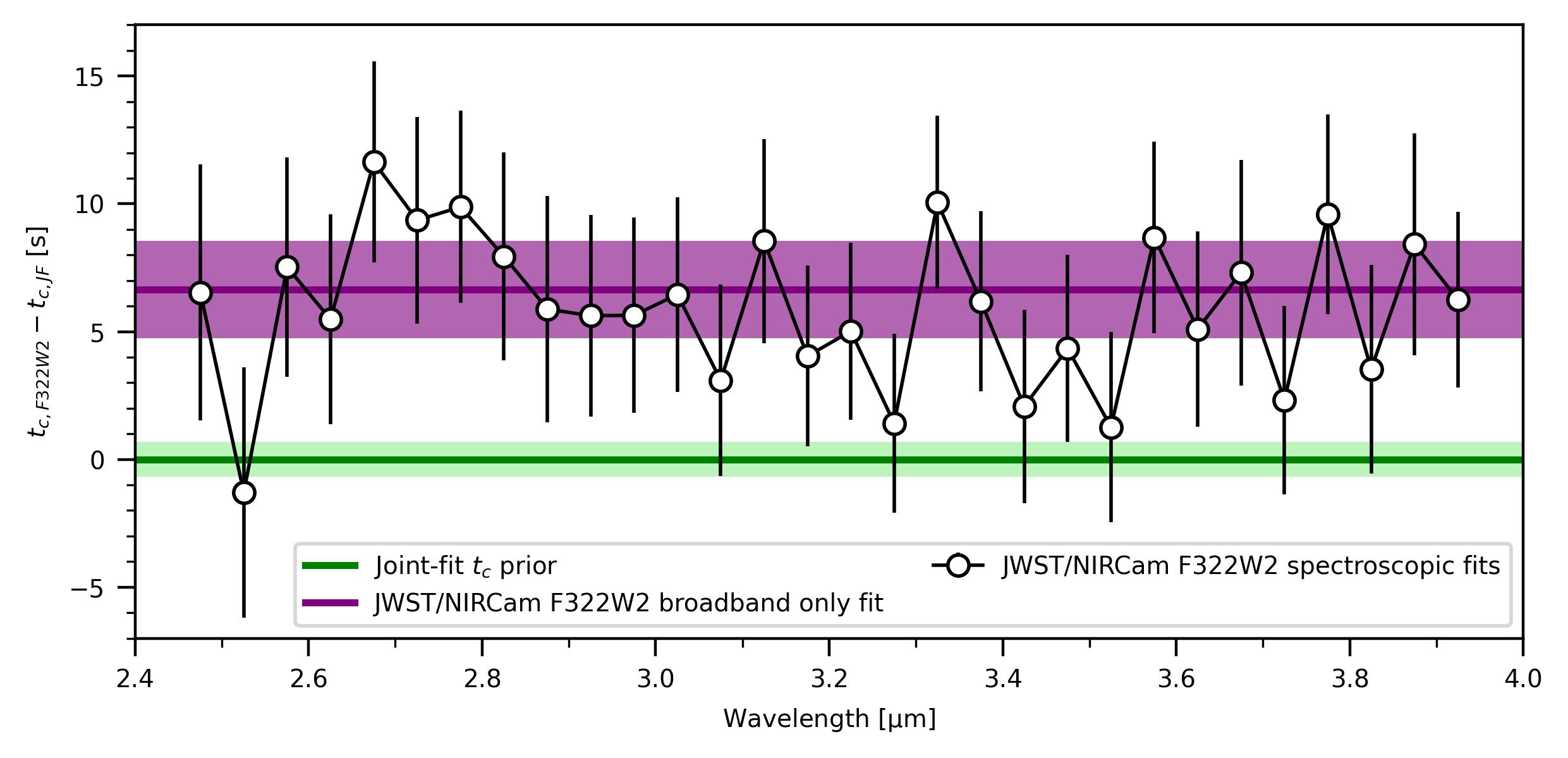}
    \caption{\textbf{Apparent chromatic offsets in WASP-107b's time of conjunction with wavelength due to limb asymmetry.} Shown are the best-fit times of conjunction as a function of wavelength when fitting our JWST/NIRCam F322W2 data by itself. We fit the data using a uniform-limb transit model, described in Section~\ref{subsec:fitting_uniformlimb}, fitting only for $t_c$, the planet-star radius ratio, and a linear flux vs. time ramp model. The black points are the result from fitting each of our thirty spectroscopic light curves. The purple line is the result from fitting the broadband light curve, with the shaded region representing the 1-$\sigma$ uncertainty. We show the difference between these results and the joint-fit time of conjunction derived by fitting other multi-wavelength data sets simultaneously, whose uncertainty is shown as the green shaded region. The observed difference is due to the significant evening-morning limb asymmetry on WASP-107b at these wavelengths.}
    \label{fig:tc_variations}
\end{figure}
\end{appendices}


\clearpage 

\section*{Data Availability}

The JWST/NIRCam (JWST GTO programme 1185; P.I. Greene; Obs. 8 and 9), JWST/MIRI (JWST GTO programme 1280; P.I. Lagage; Obs. 1), and TESS (P.I. Caldwell; Obs. I.D. hlsp\_tess-spoc\_tess\_phot\_0000000429302040-s0010\_tess\_v1) data are publicly available from the Mikulski Archive for Space Telescopes (MAST; \url{https://mast.stsci.edu}). The Spitzer/IRAC (Program 13052; P.I. Werner; AORKEY 62712320) data is publicly available from the Spitzer Heritage Archive (\url{https://irsa.ipac.caltech.edu/applications/Spitzer/SHA/}). The SOAR (Program N23A-840705; P.I. Murphy) data is available from Zenodo (\url{https://zenodo.org/records/12747273}).

\section*{Code Availability}
The three reduction pipelines used in this work (\texttt{Eureka!}, \texttt{t-shirt}, and \texttt{Pegasus}) are all either currently open-source or are planned to be made open-source. The codes of each are hosted on Github, and links for each may be found in the Methods section. 

\section*{Acknowledgements}

The authors would like to acknowledge N\'{e}stor Espinoza and Karl Misselt for assistance in verifying the reliability of JWST's instrument time-stamps, Pierre-Olivier Lagage for providing the JWST/MIRI LRS data observed in JWST program 1280, Charles Beichman for suggesting we observe this planet as part of our JWST guaranteed time program, Kevin Hardegree-Ullman for help with using the package \texttt{TTVFaster}, and N\'{e}stor Espinoza for helpful discussion about this manuscript. T.P.G. and T.J.B. acknowledge funding from NASA in WBS 411672.07.05.05.03.02. M.M., E. S., and M.R. acknowledge funding from NASA Goddard Spaceflight Center via NASA contract NAS5-02105. This work benefited from the 2023 Exoplanet Summer Program in the Other Worlds Laboratory (OWL) at the University of California, Santa Cruz, a program funded by the Heising-Simons Foundation and NASA.

\section*{Author Contributions Statement}

M.M. led this analysis including formulating the idea, performing the data and supporting analyses, and was the primary author of the paper.
T.G.B. contributed the \texttt{Pegasus} reduction of the NIRCam data, E.S. contributed the \texttt{t-shirt} reduction, and T.J.B. contributed the \texttt{Eureka!} reduction of the NIRCam F322W2 spectra and the broadband reduction of the MIRI/LRS data. T.J.B., E.S., and T.G.B. also contributed to the interpretation of our results and the writing of the paper.
M.L. contributed the modeling analysis using \texttt{ScCHIMERA}.
T.P.G. selected the object, devised the observing strategy, and contributed to the analysis. 
V.P., E.R., J.F., M.L., and L.W. each contributed to the methodology and interpretation of the results.
M.R. designed the JWST/NIRCam instrument used for the observations, and provided the observing time necessary for completing these observations.

\section*{Competing Interests Statement}

The authors declare no competing interests.

\clearpage

\clearpage 


\end{document}